\documentclass{iopart}
\input epsfig.sty

\begin{document}
\jl{4}
\topical[Nucleon-nucleon interaction]{The nucleon-nucleon interaction}
\author{R. Machleidt\dag\ and I. Slaus\ddag}
\address{\dag\ Department of Physics, University of Idaho, Moscow,
Idaho 83844, U. S. A.}
\address{\ddag\ 
 Triangle Universities Nuclear Laboratory (TUNL), Duke Station, Durham,
 North Carolina 27706, U. S. A. and
 Rudjer Boskovic Institute, Zagreb, Croatia}

\begin{abstract}
We review the major progress of the past decade
concerning our understanding of the nucleon-nucleon interaction.
The focus is on the low-energy region
(below pion production threshold),
but a brief outlook towards higher energies is also
given. The items discussed include charge-dependence,
the precise value of the $\pi NN$ coupling constant,
phase shift analysis and high-precision NN data and
potentials.
We also address the issue of a proper theory of nuclear
forces.
Finally, we summarize the essential open questions that
future research should be devoted to.
\end{abstract}

\submitted


\section{Introduction}

The nuclear force has been at the heart of nuclear physics
ever since  the field was
born in 1932 with the discovery of the neutron by Chadwick~\cite{Cha32}.
In fact, during the first few decades of nuclear physics, the term `nuclear 
forces' was often used as synonymous for 
nuclear physics as a whole~\cite{Ros48}.
There are good reasons why the nuclear force plays 
such an outstanding role.

The interaction between two nucleons is basic for all of nuclear 
physics.
The traditional goal of nuclear physics is to understand the 
properties of atomic nuclei in terms of the `bare' interaction between pairs 
of nucleons. With the onset of quantum-chromodynamics (QCD), 
it became clear that the nucleon-nucleon (NN) interaction is not fundamental.
Nevertheless, even today, in any first approach towards 
a nuclear structure problem, 
one assumes the nucleons to be elementary particles. The failure or success of 
this approach
may then teach us something about the relevance 
of subnuclear degrees of freedom.

The NN interaction has been investigated 
by a large number of physicists all over the world for the past 70 years.
It is the empirically best known piece of strong interactions; in fact, for no 
other sample of the strong force 
a comparable amount of experimental data has been accumulated.

The oldest attempt to explain  the nature of the nuclear force is
due to Yukawa~\cite{Yuk35}. According to his theory, 
 massive bosons (mesons) mediate the interaction between two nucleons.
This idea spawned the sister discipline of particle physics. 
Although, in the light of QCD, meson theory is not perceived
as fundamental anymore,
the meson exchange concept continues to represent the best working
model for a quantitative nucleon-nucleon potential.

Historically, it turned out to be a formidable task to describe
the nuclear force just phenomenologically, and it took a quarter
century to come up with the first semi-quantitative model~\cite{GT57}---in 1957. 
Ever since, there has been substantial progress in experiment and theory
of the nuclear force. Most basic questions were settled in the 1960's
and 70's such that in recent years we could concentrate
on the subtleties of this peculiar force.

In this topical review, we will report the chief progress of 
the past decade. The focus will be on the low-energy
region (below pion production threshold). Summaries of
earlier periods and a pedagogical introduction into the field
can be found in references~\cite{Mac89,ML94}.
In the 1990's, major issues concerning the NN interaction have been:
\begin{itemize}
\item
charge-dependence,
\item
the precise value of the $\pi NN$ coupling constant,
\item
improved phase shift analysis,
\item
high-precision NN data,
\item
high-precision NN potentials,
\item
QCD and the nuclear force,
\item
NN scattering at intermediate and high energies.
\end{itemize}
We will now review these topics one by one.

\section{Charge dependence}

By definition, {\it charge independence} is invariance under any 
rotation in isospin space. 
A violation of this symmetry is referred to as charge dependence
or charge independence breaking (CIB).
{\it Charge symmetry} is invariance under a rotation by 180$^0$ about the
$y$-axis in isospin space if the positive $z$-direction is associated
with the positive charge.
The violation of this symmetry is known as charge symmetry breaking (CSB).
Obviously, CSB is a special case of charge dependence.

CIB of the strong NN interaction means that,
in the isospin $T=1$ state, the
proton-proton ($T_z=+1$), 
neutron-proton ($T_z=0$),
or neutron-neutron ($T_z=-1$)
interactions are (slightly) different,
after electromagnetic effects have been removed.
CSB of the NN interaction refers to a difference
between proton-proton ($pp$) and neutron-neutron ($nn$)
interactions, only. 
The charge dependence of the NN interaction is subtle, 
but in the $^1S_0$ state it is well established. 
The observation of small charge-dependent effects in this state
is possible because the scattering length of an almost bound
state acts like a powerful magnifying glass on the interaction.

The current understanding is 
that---on a fundamental level---the 
charge dependence of nuclear forces is due to
a difference between the up and down quark masses and electromagnetic 
interactions among the quarks. 
A consequence of this
are mass differences between hadrons of the same
isospin multiplet and meson mixing.
Therefore, if CIB is calculated based upon hadronic models,
the mass differences between hadrons of the same
isospin multiplet, meson mixing, and irreducible meson-photon exchanges
are considered as major causes.
For reviews on charge dependence, see references~\cite{SAT89,MNS90,MO95}.
We will now summarize recent developments (that are not contained in
any of these reviews).

\subsection{Charge symmetry breaking}

\subsubsection{Experiment.}

As discussed, the scattering lengths in the $^1S_0$ state
for $pp$, $np$, and $nn$ scattering (denoted by $a_{pp}$,
$a_{np}$, and $a_{nn}$, respectively)
are the best evidence for the charge-dependence
of nuclear forces.
While we have well-established values for $a_{pp}$ and $a_{np}$
since many decades, the neutron-neutron scattering length
continues to be a tough problem.
The basic reason for this is that, so far,
we have not been able to conduct any direct measurements of $a_{nn}$ using
free neutron-neutron collisions~\cite{Glash}.
All current values are extracted from multi-particle reactions
the analyses of which are beset with large theoretical uncertainties.
The processes that are believed to have the smallest uncertainties
are
\begin{eqnarray}
\mu^- + d & \rightarrow & \nu_\mu + n + n \, , \\
\pi^- + d & \rightarrow & \gamma + n + n \, ,  \\
n + d & \rightarrow & p + n + n \, .
\end{eqnarray}
While, there are no data on the first reaction, the other two processes have
been studied repeatedly.
In 1998, a very carefull study of the $\pi^-$  induced reaction
was published~\cite{How98} and, in 1999, a renewed thorough investigation
of the neutron induced process was accomplished~\cite{Gon99}, yielding
results that are in perfect agreement, namely,
\begin{eqnarray}
D(\pi^-, n \gamma)n~\cite{How98}: & &
a_{nn}  =  - 18.50 \pm 0.53 \mbox{ fm,}
\\
D(n,nnp)~\cite{Gon99}: & &
a_{nn}  =  - 18.7 \pm 0.6 \mbox{ fm,}
\end{eqnarray}
which can be summarised by
\begin{equation}
a_{nn} = - 18.6 \pm 0.4 \mbox{\rm fm.}
\end{equation}
Correcting for the neutron-neutron magnetic interaction,
the pure nuclear value is:
\begin{equation}
a_{nn}^N = - 18.9 \pm 0.4 \mbox{\rm fm.}
\label{eq_ann}
\end{equation}
This summarizes the status by the end of 1999.
Unfortunately, this is not the happy end of the story
that everybody had hoped for.
To properly discuss the new (and old) problems, we will first
provide more details concerning the two
types of reactions for which experiments have been conducted.

Over the past 20 years,
there have been three independent studies of the reaction 
$ \pi^- + d \rightarrow \gamma + n + n$.
In one case~\cite{Gab79}, only the $\gamma$ spectrum was measured, 
while in the other two cases~\cite{Sch87,How98},
kinematically complete experiments were performed 
measuring the $\gamma$ and a neutron in the final state.
The results are:
\begin{eqnarray}
a_{nn} & = &
- 18.60 \pm 0.34 \mbox{ (stat.)} \pm 0.26 \mbox{ (syst.)} 
\pm 0.30 \mbox{ (theor.) fm} \nonumber \\
& = & -18.60 \pm 0.52 \mbox{ fm~\cite{Gab79},}
\\
a_{nn} & = &
- 18.70 \pm 0.42 \mbox{ (stat.)} \pm 0.39 \mbox{ (syst.)} 
\pm 0.30 \mbox{ (theor.) fm} \nonumber \\
& = & -18.70 \pm 0.65 \mbox{ fm~\cite{Sch87},}
\\
a_{nn} & = &
- 18.50 \pm 0.05 \mbox{ (stat.)} \pm 0.44 \mbox{ (syst.)} 
\pm 0.30 \mbox{ (theor.) fm} \nonumber \\
& = & -18.50 \pm 0.53 \mbox{ fm~\cite{How98}.}
\end{eqnarray}
	Owing to the high spatial resolution of the gamma ray detector in
reference~\cite{How98}, it was possible to assess the systematic errors due to
uncertainties in the modelling of the stopped pion distribution in the
target and in target vertex reconstruction in the Monte Carlo
simulation. Therefore, the systematic uncertainties of the kinematically
complete studies are now much better understood, and a very high
statistical accuracy in reference~\cite{How98} makes the experimental uncertainty
comparable to the theoretical one in the extraction of $a_{nn}$ from
the reaction D$(\pi^-, \gamma n)n$.
The combined result from all three studies gives the new world
average for the D$(\pi^-,\gamma)nn$ reaction
\begin{eqnarray}
a_{nn} & = & -18.59 \pm 0.27 \mbox{ (exper.)} \pm 0.30 \mbox{ (theor.) fm}
\nonumber \\
& = & -18.59 \pm 0.40 \mbox{ fm.}
\label{eq_annav}
\end{eqnarray}
In summary, the reaction $\pi^-+d\rightarrow \gamma+n+n$ appears to be
in good shape. 

Unfortunately, we cannot say the same about the neutron-induced
deuteron breakup process. 
Until the recent investigation by the TUNL group,
Gonzalez-Trotter {\it et al.}~\cite{Gon99}, all studies of
the reaction 
$n+d\rightarrow p+n+n$
gave for $a_{nn}$ 
values that differed from that obtained
from the D$(\pi^-, \gamma n)n$ process. Theoretical uncertainties
in extracting $a_{nn}$ from the neutron-induced deuteron breakup are much
larger, as we will explain now.

	First, in reactions with more than two nucleons in the
final state, three nucleon
forces (3NF) modify the cross section. It was suggested~\cite{Sla82}
that the 3NF is the reason why $a_{nn}$ extracted from
the D$(n,nnp)$ process differs from that obtained from the 
D$(\pi^-, \gamma n)n$ reaction. 
The 3NF are a natural consequence of strong
interactions. Therefore, 3NF do exist, but the question is how significant
they are and, in particular, do they affect a specific configuration of the
$nd$ breakup wherefrom one extracts $a_{nn}$. There are several indications for
possible 3NF effects in nuclear physics: 
$^3$H binding energy, nuclear matter binding energy, $^4$He
binding energy and negative parity excited states, $^3$He and $^4$He one-body
density distributions, binding energies and radii of some nuclei,
$^{17}$O magnetic moment form factor, $nd$ capture, $A_y$ in elastic $nd$
scattering, and space star, final state interaction (FSI) and quasifree
scattering (QFS) configurations in the $nd$ breakup~\cite{SAT89,Glo96,Set96}; 
but none of them provided conclusive information on 3NF. 
It was possible to reconcile all
values of $a_{nn}$ extracted from the studies of the reaction D$(n,nnp)$ in the
energy domain of 10 to 50 MeV with those obtained from the 
D$(\pi^-, \gamma n)n$ 
process using the Fujita-Miyazawa 3NF~\cite{SAT89}.
However, the reanalyses of these $nd$ breakup
processes~\cite{Tor96} 
gave values that differed considerably from those quoted by the original
authors. Though any reanalysis is clouded by the lack of all relevant
information, the main reason is the fact that original analyses used
simple $S$-wave separable NN potentials, while the re-analyses were done
using the rigorous three body theory of Gl\"ockle {\it et al}~\cite{Glo96}.
Obviously, the claimed theoretical
uncertainties in the original papers were underestimated.

	Second, the magnetic interaction modifies the value of the $^1S_0$
scattering length extracted from the neutron induced deuteron breakup.  
It was shown~\cite{Slo84}
that for the neutron-pickup
configuration in the neutron-induced deuteron breakup 
leading to the $nn$ FSI there
is a magnetic interaction in the $^1S_0$ state which is repulsive thereby
decreasing the absolute value of $a_{nn}$. Depending on the NN potential (hard
core or soft core), impulse approximation estimates of the effect of the
magnetic interaction in the pickup configuration changes $a_{nn}$ 
from $-18.5$ fm
to $-17.2$ or even $-16.4$ fm~\cite{Slo84}. The correction for the knockout
configuration has the opposite sign since it is dominated by the magnetic
interaction between a neutron and a proton in the $^1S_0$ state. 
The situation is more
complex for the neutron-proton FSI, since the $np$ FSI occurs in the $^1S_0$ 
and $^3S_1$ states.

	The determination of $a_{nn}$ by 
Gonzales-Trotter {\it et al}~\cite{Gon99} has two
characteristic features: first, it uses the rigorous theory~\cite{Glo96}
including, in addition to several realistic NN potentials, also the
Tucson-Melbourne 3NF, and second, it performs a high-accuracy
comparison of neutron-neutron and neutron-proton FSI in the $^1S_0$
state by measuring cross sections of the reaction D$(n,nnp)$ for identical
kinematic conditions (the angle of the two emitted nucleons interacting in the
$^1S_0$ final state is 28 to 43 deg) at the incident neutron energy of 13
MeV. Therefore, the neutron-proton scattering length, $a_{np}$, becomes the
standard for determining $a_{nn}$. By comparing the extracted value for 
$a_{np}$ and its uncertainty, 
it was possible to set an upper limit of 0.2 $\pm$ 0.6 fm on
any possible effects due to 3NF influencing the extracted value of $a_{nn}$. 
Of course, it is possible that the effect of the magnetic interaction
discussed by Slobodrian {\it et al}~\cite{Slo84}
 are negligible in the energy/angular region
studied by Gonzalez-Trotter {\it et al}~\cite{Gon99}. 
On the other hand, it should be stressed that
the rigorous calculations by Gl\"ockle {\it et al}~\cite{Glo96}
do not include the electromagnetic interaction, and
that there are now many indications of the shortcomings of the 
Tucson-Melbourne 3NF.  

In the year of 2000, a new study of
the neutron-neutron FSI in the D$(n,nnp)$ reaction 
at the incident neutron energy of 25.3 MeV
was published by the neutron group at Bonn~\cite{Huh00}.
The data were analyzed using the rigorous theory~\cite{Glo96}. 
The extracted value is,
\begin{equation}
a_{nn} = - 16.27 \pm 0.4 \mbox{ fm,}
\end{equation}
which is
in drastic disagreement with the result of the TUNL group~\cite{Gon99} 
published in 1999 and also with
those obtained from the D$(\pi^-, \gamma n)n$ process. 

	While most of the previous kinematically complete studies of the
reaction D$(n,nnp)$ employ a thick, active deuterated target measuring the
energy of the proton, and detecting two neutrons at nearly the same
angle on the same side of the incident neutron, 
this recent measurement~\cite{Huh00}
uses a thin deuterium target and detects a neutron 
at $\Theta_n = -55.5$ deg 
and a proton at $\Theta_p = 41.15$ deg. 
The advantage of
this geometry is the reduction of the strong cross talk between neutron
detectors and the reduction in losses from neutron multiple scattering.  
This geometrical configuration has the added advantage that the locus
contains $np$ QFS besides $nn$ FSI and, therefore, provides a built-in
normalization. Indeed, normalizing the data to $np$ QFS yields a very
similar value:
\begin{equation}
a_{nn} = - 16.06 \pm 0.35 \mbox{ fm.}
\end{equation}
Neither the use of different NN potentials nor the inclusion of the
Tucson-Melbourne 3NF in the rigorous calculation produces noticeably
different results for $a_{nn}$. This geometry at this incident energy is the
region where the 3NF effect of the Tucson-Melbourne potential is very
small. The preliminary result by the Bonn group---using the same
incident energy of 25.3 MeV---gave a good fit to the $np$ FSI spectrum using 
$a_{np} = - 24$ fm.

	The disagreement between the two most recent studies, 
Gonzalez-Trotter {\it et al} (TUNL)~\cite{Gon99} 
and Huhn {\it et al} (Bonn)~\cite{Huh00},
opens again the problem of how completely do we understand the
interactions involved in the three nucleon problem, specifically the 3NF.   
It also suggests that additional experimental studies at different incident
energies and at different angles might be useful in resolving the problem.

When we use for $a_{nn}$ the value obtained from the
D$(\pi^-,\gamma)nn$ studies
[equation~(\ref{eq_annav})],
correct it for the magnetic moment interaction [equation~(\ref{eq_ann})],
and compare it to the corresponding $pp$ value~\cite{MNS90}:
\begin{equation}
a^N_{pp}=-17.3\pm 0.4 \mbox{ fm,} 
\end{equation}
then
charge-symmetry is broken by the following amount,
\begin{equation}
\Delta a_{CSB} \equiv a_{pp}^N-a_{nn}^N = 1.6\pm 0.6 \mbox{ fm}. 
\label{eq_CSBlep1}
\end{equation}
Recommended values for the corresponding effective ranges are~\cite{MNS90},
\begin{eqnarray}
r^N_{nn} & = & 2.75 \pm 0.11 \mbox{ fm,}
\\
r^N_{pp} & = & 2.85 \pm 0.04 \mbox{ fm,}
\end{eqnarray}
implying
\begin{equation}
\Delta r_{CSB} \equiv r_{pp}^N-r_{nn}^N = 0.10\pm 0.12 \mbox{ fm}. 
\label{eq_CSBlep2}
\end{equation}

	Traditionally, it was believed that the meson mixing explains
essentially all CSB effects. The largest contribution came from
$\rho-\omega$ mixing, and there was very meager knowledge of 
$\pi-\eta$ mixing.
Recently, CSB was studied by the comparison of two charge-symmetric
processes:
D$(\pi^+,\eta)pp$ and D$(\pi^-, \eta)nn$ in the energy region of the $\eta$
threshold. The result for the ratio of the two processes in this energy
region is $R = d\sigma^+/d\sigma^-  = 0.937 \pm 0.007$, after a phase space
correction is made for the difference in the threshold energies of the two
reactions~\cite{Tip00}.
The deviation of $R$ from 1 is an indication of CSB which is mostly due to
$\pi-\eta$ mixing. A phenomenological fully relativistic model,
which is based on coupled channel N$\pi$--N$\eta$ amplitudes, 
takes into account different $nn$ and $pp$ FSI and explicitly includes
$\pi-\eta$ mixing, 
was developed~\cite{Bat98}
and compared to the data yielding for the
$\pi-\eta$ mixing angle the value of $(1.5 \pm 0.4)$ deg, 
consistent with the mixing
angle determined from particle decays and isospin-forbidden processes as
well as with several other theoretical predictions~\cite{Tip00}.

\subsubsection{Theory.}
The difference between the masses of  neutron and proton
represents the most basic cause for CSB of the nuclear force. 
Therefore, it is important to have a very thorough 
accounting of this effect. 

The most trivial consequence of nucleon mass splitting is a difference
in the kinetic energies: for the heavier neutrons, the kinetic energy
is smaller than for protons. This raises the magnitude of the
$nn$ scattering length by 0.25 fm as compared to $pp$.
 
Besides the above, nucleon mass splitting has an impact on all meson-exchange
diagrams that contribute to the nuclear force.
In 1998, the most comprehensive and thorough calculation of these CSB effects
ever conducted has been published~\cite{LM98a}.
The investigation is based upon the Bonn Full Model for the 
NN interaction~\cite{MHE87}.
Here, we will summarize the results.
For this we devide the total number of meson exchange diagrams that
is involved in the nuclear force into several classes. 
Below, we report the results for each class.

\begin{figure}
\vspace*{-3cm}
\hspace*{0.0cm}
\epsfig{file=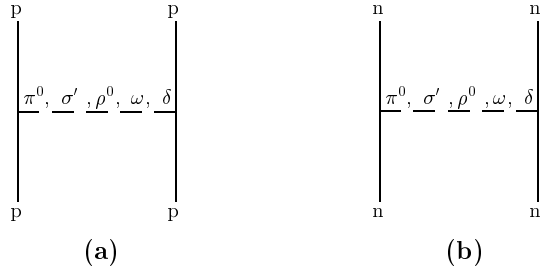,width=12cm}

\vspace*{-9cm}
\caption{One-boson-exchange (OBE) contributions to (a) $pp$
and (b) $nn$ scattering.}
\label{fig_csb1}
\end{figure}

\begin{table}
\caption{CSB differences of the $^1S_0$ effective range parameters
caused by nucleon mass splitting.
$2\pi$ denotes the sum of all $2\pi$-contributions and $\pi\rho$
the sum of all $\pi\rho$-contributions. TBE (non-iterative two-boson-exchange)
is the sum of $2\pi$, $\pi\rho$, and $(\pi\sigma+\pi\omega)$.
}
\footnotesize
\begin{tabular}{lcccccccc}
\br
               &kin.\ en.\ & OBE &$2\pi$ &
 $\pi\rho$ &
 $\pi\sigma+\pi\omega$& TBE & Total & Empirical \\
\mr 
$\Delta a_{CSB}$ (fm) &
  0.246 & 0.013 & 2.888 & --1.537 & --0.034 & 1.316 & 1.575 & $1.6\pm 0.6$ fm\\
$\Delta r_{CSB}$ (fm) &
  0.004 & 0.001 & 0.055 & --0.031 & --0.001 & 0.023 & 0.028 & $0.10\pm 0.12$ fm
\\
\br
\end{tabular}
\label{tab_csb}
\end{table}

\begin{figure}
\vspace*{-1.5cm}
\hspace*{0.0cm}
\epsfig{file=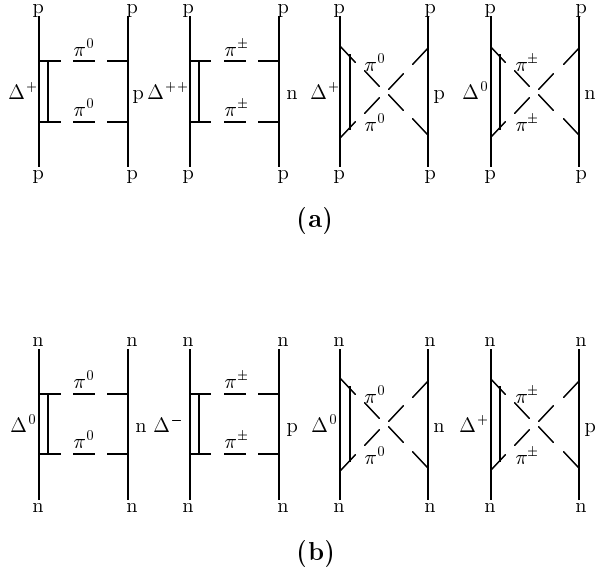,width=12cm}

\vspace*{-6.5cm}
\caption{Two-pion-exchange contributions with $N\Delta$ intermediate
states to (a) $pp$ and (b) $nn$ scattering.}
\label{fig_csb2}
\end{figure}

\begin{figure}
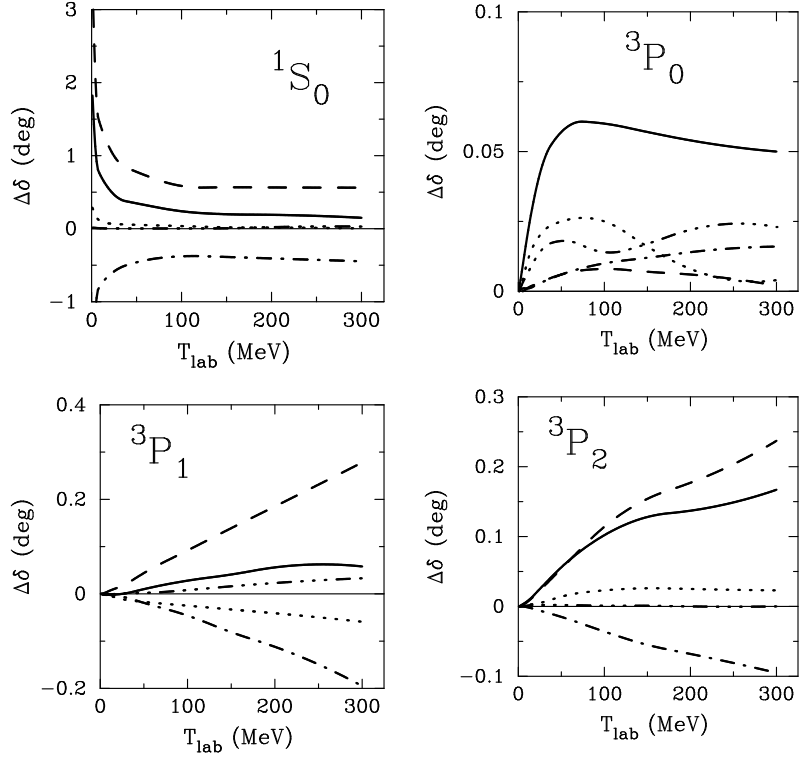


\vspace{1.0cm}
\hspace*{2.25cm}
\epsfig{file=fig_csb3.ps,width=5cm}

\vspace*{-4.9cm}
\hspace*{7.75cm}
\epsfig{file=fig_csb4.ps,width=5cm}

\vspace*{0.50cm}
\hspace*{2.25cm}
\epsfig{file=fig_csb5.ps,width=5cm}

\vspace*{-4.9cm}
\hspace*{7.75cm}
\epsfig{file=fig_csb6.ps,width=5cm}

\vspace*{0.5cm}
\caption{CSB phase shift differences $\delta_{nn}-\delta_{pp}$ (without
electromagnetic interactions)
for laboratory kinetic energies $T_{lab}$
below 300 MeV and partial waves with
$L\leq 1$.
The CSB effects due to the kinetic energy, OBE, the entire $2\pi$ model,
and $\pi\rho$ exchanges
are shown by the dotted, dash-triple-dot, dashed, and dash-dot
curves, respectively.
The solid curve is the sum of all CSB effects.}
\label{fig_csb3}
\end{figure}

\begin{enumerate}
\item
{\bf One-boson-exchange} (OBE, figure~\ref{fig_csb1}) contributions
mediated by $\pi^{0}(135)$, $\rho^0(770)$,
 $\omega(782)$, $a_0/\delta(980)$, and
$\sigma'(550)$. 
In the Bonn Full Model~\cite{MHE87},
the $\sigma'$ describes only the correlated $2\pi$ exchange in 
$\pi\pi-S$-wave (and not the uncorrelated $2\pi$ exchange
since the latter is calculated explicitly, cf.\ figure~\ref{fig_csb2}).
Charge-symmetry is broken by the fact that
for $pp$ scattering the proton mass is used in the Dirac spinors
representing the four external legs 
[figure~\ref{fig_csb1}(a)], 
while for $nn$ scattering the neutron mass
is applied 
[figure~\ref{fig_csb1}(b)]. 
The CSB effect from the OBE diagrams is very small
(cf.\ table~\ref{tab_csb}).

\item
{\bf $2\pi$-exchange diagrams.} 
This class consists of three groups;
namely the diagrams with NN, N$\Delta$ and $\Delta\Delta$
intermediate states, where $\Delta$ refers to the baryon with
spin and isospin $\frac32$ and mass 1232 MeV.
The most important group is the one with $N\Delta$ intermediate
states which we show in figure~\ref{fig_csb2}.
Part (a) of figure~2 applies to $pp$ scattering, while part (b)
refers to $nn$ scattering.
When charged-pion exchange is involved, the intermediate-state
nucleon differs from that of the external legs. This is one of the
sources for CSB from this group of diagrams.
The $2\pi$ class of diagrams causes the largest CSB effect
(cf.\ table~\ref{tab_csb} and dashed curve in figure~\ref{fig_csb3}).

\item
{\bf $\pi\rho$-exchanges.} 
Graphically, the $\pi\rho$ diagrams can be obtained
by replacing in each $2\pi$ diagram (e.~g., in figure~\ref{fig_csb2})
one pion by a $\rho$-meson of the same charge state.
The effect is typically opposite to the one from $2\pi$ exchange.

\item
{\bf Further $3\pi$ and $4\pi$ contributions} ($\pi\sigma+\pi\omega$).
The Bonn potential also includes some $3\pi$-exchanges that can be
approximated in terms of $\pi\sigma$ diagrams and $4\pi$-exchanges
of  $\pi\omega$ type.
The sum of the two groups is small, indicating convergence of the 
diagramatic expansion. 
The CSB effect from this class is essentially negligible.
\end{enumerate}

The total CSB difference of the singlet scattering length 
caused by nucleon mass splitting amounts
to 1.58 fm (cf.\  table~\ref{tab_csb}) which agrees well with
the empirical value $1.6\pm 0.6$ fm.
Thus, nucleon mass splitting alone can explain the entire
empirical CSB of the singlet scattering length~\cite{CN96}.
{\it This is a remarkable result.}

The impact of the various classes of diagrams on CSB phase shift
differences are shown in figure~\ref{fig_csb3}.
The total effect is the largest in the $^1S_0$ state
where it is most noticable at low energy; e.~g., at 1 MeV, 
the phase shift difference is 1.8 deg.
The difference decreases with increasing energy
and is about 0.15 deg at 300 MeV, in $^1S_0$.

The CSB effect on the phase shifts of
higher partial waves is small;
in $P$ and $D$ waves, typically in the order of 0.1 deg 
at 300 MeV and less at lower energies.
This fact may suggest that CSB in partial waves other than
$L=0$ may be of no relevance. 
In references~\cite{Mut99} it was shown that this is not true:
CSB beyond the $S$ waves is crucial for the explanation of
the Nolen-Schiffer anomaly.

Before finishing this subsection,
a word is in place concerning other mechanisms that cause
CSB of the nuclear force.
Traditionally, it was believed that 
$\rho^0-\omega$ 
mixing explains essentially all CSB in the nuclear force~\cite{MNS90}.
However, recently some doubt has been cast on this paradigm.
Some researchers~\cite{GHT92,PW93,KTW93,Con94} found that 
$\rho^0-\omega$ exchange may have a substantial
$q^2$ dependence such as to cause this contribution to nearly vanish
in NN.
Our finding that the empirically known CSB in the 
nuclear force can be explained solely from nucleon mass splitting 
(leaving essentially no room for additional CSB contributions 
from $\rho^0-\omega$ mixing or other sources) fits well into this 
scenario.
On the other hand, Miller~\cite{MO95} and Coon and coworkers~\cite{CMR97}
have advanced counter-arguments that would restore the traditional role
of $\rho$-$\omega$ exchange.
The issue is unresolved.
Good summaries of the controversial points of view can be found in
references~\cite{MO95,Con97,CS00}.

Finally, for reasons of completeness, we mention that irreducible
diagrams of $\pi$ and $\gamma$ exchange between
two nucleons create a charge-dependent nuclear force.
Recently, these contributions have been calculated to
leading order in chiral perturbation theory~\cite{Kol98}.
It turns out that to this order the $\pi\gamma$ force is
charge-symmetric (but does break charge independence).

\subsection{Charge independence breaking}

\begin{figure}[t]

\vspace*{-3.0cm}
\hspace*{0.0cm}
\epsfig{file=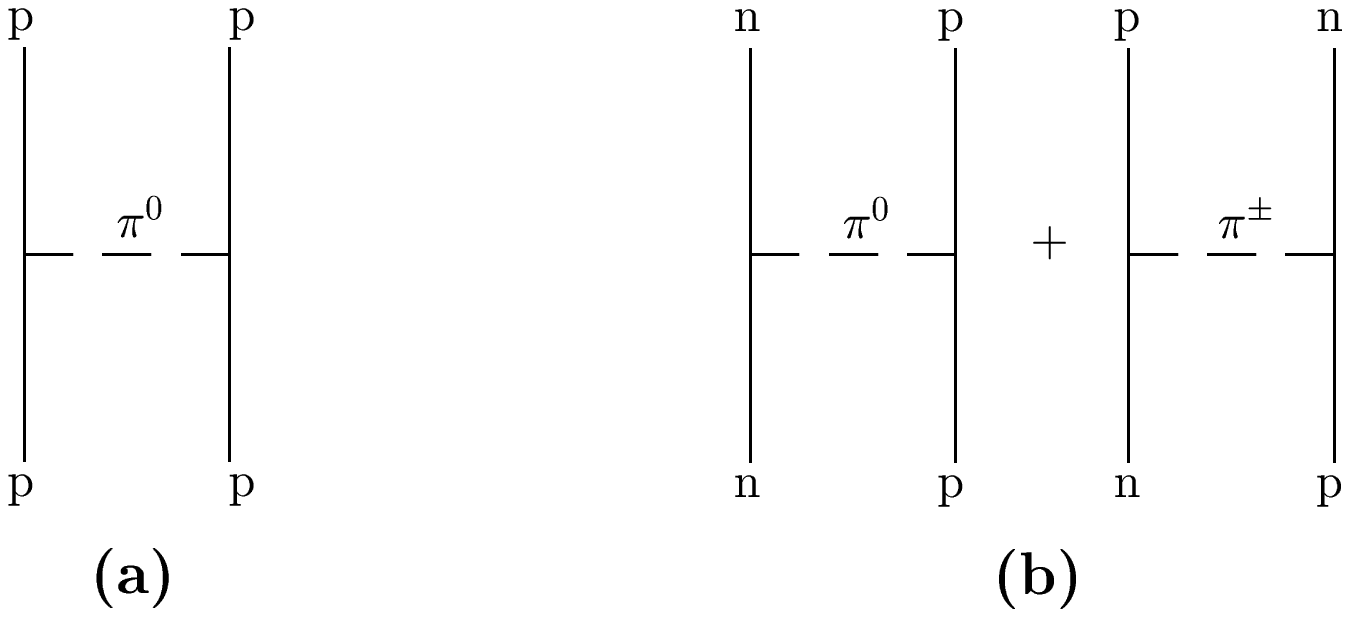,width=12cm}

\vspace*{-9.0cm}
\caption{One-pion exchange (OPE) contributions to
(a) $pp$ and (b) $np$ scattering.} 
\label{fig_cib1}
\end{figure}

The empirical values for the $np$ singlet 
effective range parameters are~\cite{KMS84}:
\begin{equation}
a_{np}=-23.740\pm 0.020 \mbox{ fm}, \hspace*{.2cm}
     r_{np}=2.77\pm 0.05 \mbox{ fm}.
\end{equation}
It is useful to define the following averages:
\begin{eqnarray}
\bar{a}\equiv \frac12 (a^N_{pp} + a^N_{nn}) & =&  -18.1\pm 0.6 \mbox{ fm},\\
\bar{r}\equiv \frac12 (r^N_{pp} + r^N_{nn}) & =&  2.80\pm 0.12 \mbox{ fm}.
\end{eqnarray}
Ignoring CSB, the CIB differences in the effective range parameters
are given by:
\begin{eqnarray}
\Delta a_{CIB} \equiv
 \bar{a}
 - 
 a_{np}
 &=& 5.64\pm 0.60 \mbox{ fm},\\
\Delta r_{CIB} \equiv
 \bar{r}
 - 
 r_{np}
 &=& 0.03\pm 0.13 \mbox{ fm}.
\end{eqnarray}

The major cause of CIB in the NN interaction is pion mass splitting.
Based upon the Bonn Full Model for the NN interaction~\cite{MHE87},
the CIB due to pion mass splitting has been calculated 
carefully and systematically in reference~\cite{LM98b}.
We will discuss now the various classes of diagrams
and their contributions to CIB.

\begin{table}[b]
\caption{CIB contributions to the $^1S_0$ scattering length,
$\Delta a_{CIB}$, 
and effective range, 
$\Delta r_{CIB}$,
from various components of the NN interaction.}
\begin{indented}
\item[]\begin{tabular}{lcccccc}
\br
   & OPE & $2\pi$  & $\pi\rho$ & $\pi\sigma+\pi\omega$ & Total&Empirical  \\
\mr 
$\Delta a_{CIB}$ (fm) &
 3.243 & 0.360 & -0.383 & 1.426 & 4.646 & $5.64\pm 0.60$
\\
$\Delta r_{CIB}$ (fm) &
 0.099 & 0.002 & -0.006 & 0.020 & 0.115 & $0.03\pm 0.13$
\\
\br
\end{tabular}
\end{indented}
\label{tab_cib}
\end{table}

\begin{figure}
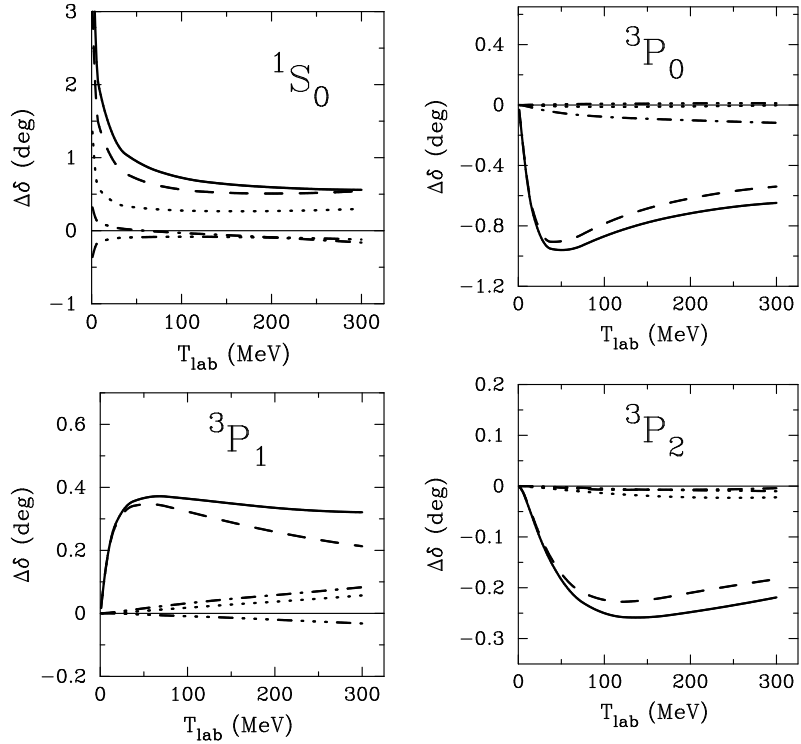


\vspace{1.0cm}
\hspace*{2.25cm}
\epsfig{file=fig_cib3.ps,width=5cm}

\vspace*{-4.9cm}
\hspace*{7.75cm}
\epsfig{file=fig_cib4.ps,width=5cm}

\vspace*{0.50cm}
\hspace*{2.25cm}
\epsfig{file=fig_cib5.ps,width=5cm}

\vspace*{-4.9cm}
\hspace*{7.75cm}
\epsfig{file=fig_cib6.ps,width=5cm}

\vspace*{0.5cm}
\caption{CIB phase shift differences $\delta_{np} - \bar{\delta}$
[with $\bar{\delta} \equiv ( \delta_{pp} + \delta_{nn})/2$]
for laboratory kinetic energies $T_{lab}$
below 300 MeV and partial waves with orbital angular momentum
$L\leq 1$.
The CIB effects due to OPE, the entire $2\pi$ model,
$\pi\rho$ exchanges, and $(\pi\sigma+\pi\omega)$ contributions
are shown by the dashed, dash-dot, dash-triple-dot, and dotted
curves, respectively.
The solid curve is the sum of all CIB effects.}
\label{fig_cib3}
\end{figure}

\begin{figure}[t]

\vspace*{-3.0cm}
\hspace*{0.0cm}
\epsfig{file=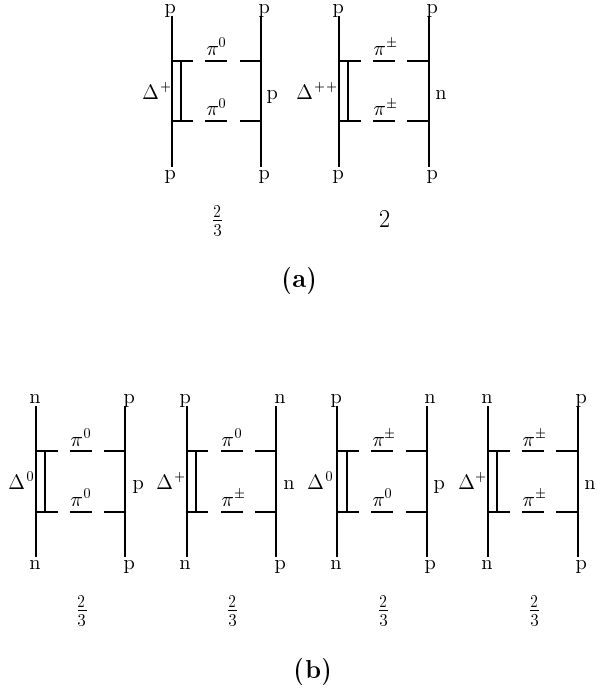,width=12cm}

\vspace*{-4.5cm}
\caption{$2\pi$-exchange box diagrams
with $N\Delta$ intermediate states that contribute to 
(a) $pp$ and (b) $np$ scattering. 
The numbers below the diagrams are
the isospin factors.}
\label{fig_cib2}
\end{figure}

\begin{enumerate}
\item
{\bf One-pion-exchange} (OPE). 
The CIB effect is created by replacing the diagram 
figure~\ref{fig_cib1}(a)
by the two diagrams 
figure~\ref{fig_cib1}(b).
The effect caused by this replacement can be understood as follows.
In nonrelativistic approximation\footnote{For pedagogical reasons, 
we use simple, approximate
expressions to discuss the effects from pion exchange.
Note, however, that in the calculations of reference~\cite{LM98b}
relativistic time-ordered perturbation theory is applied
in its full complexity and without approximations.}
and disregarding isospin factors, OPE is given by
\begin{equation}
V_{1\pi}(g_\pi, m_\pi)  =  -\frac{g_{\pi}^{2}}{4M^{2}}
 \frac{({\mbox {\boldmath $\sigma$}}_{1} \cdot {\bf k})
       ({\mbox {\boldmath $\sigma$}}_{2} \cdot {\bf k})}
{m_{\pi}^2+{\bf k}^{2}} \;
F^2_{\pi NN} (\Lambda_{\pi NN},|{\bf k}|)
\label{eq_ope}
\end{equation}
with $M$ the average nucleon mass, $m_\pi$ the pion mass, 
and {\bf k} the momentum transfer.
The above expression includes a $\pi NN$ vertex form-factor,
$F_{\pi NN}$, which depends on the
cutoff mass $\Lambda_{\pi NN}$ and the magnitude of the momentum transfer
$|{\bf k}|$.
For $S=0$ and $T=1$, where $S$ and $T$ denote the total spin and isospin
of the two-nucleon system,
respectively, we have
\begin{equation}
^{01}V_{1\pi}(g_\pi, m_\pi)  =  
\frac{g_{\pi}^{2}}
{m_{\pi}^2+{\bf k}^{2}}
\;
\frac{{\bf k}^2}
{4M^{2}} \;
F^2_{\pi NN} (\Lambda_{\pi NN},|{\bf k}|)
\, ,
\label{eq_pi1s0}
\end{equation}
where the superscripts 01 refer to $ST$.
In the $^1S_0$ state, this potential expression is repulsive.
The charge-dependent OPE is then,
\begin{equation}
^{01}V_{1\pi}^{pp}  =  
\: ^{01}V_{1\pi}(g_{\pi^0}, m_{\pi^0})  
\end{equation}
for $pp$ scattering, and
\begin{equation}
^{01}V_{1\pi}^{np}  =  
2\: ^{01}V_{1\pi}(g_{\pi^\pm}, m_{\pi^\pm})  
-\: ^{01}V_{1\pi}(g_{\pi^0}, m_{\pi^0})  
\end{equation}
for $np$ scattering.
If we assume charge-independence of $g_\pi$ (i.~e., 
$g_{\pi^0}=g_{\pi^\pm}$), then all CIB comes from the charge
splitting of the pion mass, which is~\cite{PDG98}
\begin{eqnarray}
m_{\pi^0} & = & 134.977 \mbox{ MeV,}\\
m_{\pi^\pm} & = & 139.570 \mbox{ MeV.}
\end{eqnarray}
Since the pion mass appears in the denominator of OPE,
the smaller $\pi^0$-mass exchanged in $pp$ scattering
generates a larger (repulsive) potential in the $^1S_0$
state as compared to $np$ where also the heavier $\pi^\pm$-mass
is involved. Moreover, the $\pi^0$-exchange in $np$
scattering carries a negative sign, 
which further weakens the $np$ OPE potential.
The bottom line is that the $pp$ potential is more repulsive
than the $np$ potential. The quantitative effect on
$\Delta a_{CIB}$
is such that it explains about 60\% of the empirical
value (cf.\ table~\ref{tab_cib}).
This has been know for a long time.

Due to the small mass of the pion, OPE is a sizable
contribution in all partial waves including higher partial waves; 
and due to the pion's relatively large mass splitting (3.4\%),
OPE creates relatively large charge-dependent effects 
in all partial waves (see dashed curve in figure~\ref{fig_cib3}).

\item
{\bf $2\pi$-exchange diagrams.}
We now turn to the CIB created by the $2\pi$ exchange contribution
to the NN interaction. There are many diagrams that
contribute (see reference~\cite{LM98b} for a complete overview).
For our qualitative discussion here, we pick the largest
of all $2\pi$ diagrams, namely, the box diagrams with
$N\Delta$ intermediate states, figure~\ref{fig_cib2}.
Disregarding isospin factors and using some drastic
approximation, the amplitude for such a diagram is
\begin{eqnarray}
\hspace*{-1.5cm}
V_{2\pi}(g_\pi, m_\pi) &=&  -\frac{g_{\pi}^{4}}{16M^{4}}
\frac{72}{25} \int \frac{d^3p}{(2\pi)^3}
 \frac{[{\mbox {\boldmath $\sigma$}} \cdot {\bf k}
        {\mbox {\boldmath $S$}} \cdot {\bf k}]^2}
{(m_{\pi}^2+{\bf k}^{2})^2(E_p+E^\Delta_p-2E_q)}
\nonumber \\
&&
\times
F^2_{\pi NN} (\Lambda_{\pi NN},|{\bf k}|) \;
F^2_{\pi N\Delta} (\Lambda_{\pi N\Delta},|{\bf k}|)
\, ,
\label{eq_2pi}
\end{eqnarray}
where ${\bf k} = {\bf p} - {\bf q}$ with {\bf q}
the relative momentum in the initial and final state
(for simplicity, we are considering a diagonal matrix element); 
$E_p=\sqrt{M^2+{\bf p}^2}$ and $E^\Delta_p=\sqrt{M_\Delta^2+{\bf p}^2}$
with $M_\Delta=1232$ MeV the $\Delta$-isobar mass;
{\bf S} is the spin transition operator between nucleon and
$\Delta$. For the $\pi N\Delta$ coupling constant, $f_{\pi N\Delta}$,
the quark-model relationship
$f^2_{\pi N\Delta} = \frac{72}{25} f^2_{\pi NN}$ 
is used~\cite{MHE87}.

For small momentum transfers {\bf k},
this attractive contribution is roughly proportional to
$m_\pi^{-4}$. Thus for the $2\pi$ exchange, 
the heavier pions will provide less attraction
than the lighter ones.
Charged and neutral pion exchanges
occur for $pp$ as well as for $np$, and it is important
to take the isospin factors carried by the various diagrams
into account. They are given in 
figure~\ref{fig_cib2} 
below each diagram.
For $pp$ scattering, the diagram with 
double $\pi^\pm$ exchange
carries the largest factor, while 
double $\pi^\pm$ exchange
carries only a small relative weight in $np$ scattering.
Consequently, $pp$ scattering is less attractive than $np$
scattering which leads to an increase of
$\Delta a_{CIB}$ by 0.79 fm due to the diagrams of 
figure~\ref{fig_cib2}. 
The crossed diagrams of this type
reduce this result and including all $2\pi$ exchange diagrams 
one finds a total effect of 0.36 fm~\cite{LM98b}.

\item
{\bf  $\pi\rho$-exchanges.} This group is, in principle, as comprehensive
as the $2\pi$-exchanges discussed above. Graphically, 
the $\pi\rho$ diagrams can be obtained
by replacing in each $2\pi$-diagram one of the two pions by a
$\rho$-meson of the same charge-state.
This contribution to CIB (dash-triple-dot curve 
in figure~\ref{fig_cib3}) is generally small, and
(in most states) opposite to the one from $2\pi$.

\item
{\bf Further $3\pi$ and $4\pi$ contributions} ($\pi\sigma+\pi\omega$).
As discussed,
the Bonn potential also includes some $3\pi$-exchanges that can be
approximated in terms of $\pi\sigma$ diagrams and $4\pi$-exchanges
of  $\pi\omega$ type.
These diagrams carry the same isospin factors as OPE.
The CIB effect from this class is very small, except in
$^1S_0$ (dotted curve in figure~\ref{fig_cib3}).
Notice that this effect has always the same sign as the 
effect from OPE (dashed curve), 
but it is substantially smaller. 
The reason for the OPE character of this contribution is that
$\pi\sigma$ prevails over $\pi\omega$ and, thus, determines the
character of this contribution. 
Since sigma-exchange is negative and since, futhermore,
the propagator in between the $\pi$ and the $\sigma$ exchange is also
negative, the overall sign of the $\pi\sigma$ exchange is the same as OPE.
Thus, it is like a short-ranged OPE contribution.
\end{enumerate}

Concerning the singlet scattering length, the CIB contributions discussed
explain about 80\% of $\Delta a_{CIB}$ (cf.\ table~\ref{tab_cib}).
Ericson and Miller~\cite{EM83} arrived at a very similar result using the
meson-exchange model of Partovi and Lomon~\cite{PL70}.

The sum of all CIB effects on phase shifts is shown by 
the solid curve in figure~\ref{fig_cib3}.
Notice that the difference between the solid curve
and the dashed curve (OPE) in that figure represents the sum of
all effects beyond OPE.
Thus, it is clearly seen that OPE dominates the CIB effect in all
partial waves, even though there are substantial contributions besides
OPE in some states, notably $^1S_0$ and $^3P_1$.

In reference~\cite{LM98b}, also the effect
of rho-mass splitting on the $^1S_0$ effective range parameters
was investigated.
Unfortunately, the evidence for rho-mass splitting is very
uncertain, with the Particle Data Group~\cite{PDG98} reporting
$m_{\rho^0} - m_{\rho^\pm} = 0.4 \pm 0.8$ MeV.
Consistent with this,  
$m_{\rho^0} = 769$ MeV and
$m_{\rho^\pm} = 768$ MeV, 
i.~e., a splitting of 1 MeV was assumed, 
in the exploratory study of reference~\cite{LM98b}.
With this, one finds 
$\Delta a_{CIB} = -0.29$ fm 
from  
one-rho-exchange, and 
$\Delta a_{CIB} = 0.28$ fm 
from the non-iterative $\pi\rho$ diagrams with NN intermediate
states. Thus, individual effects are small and, in addition,
there are substantial cancellations between
the two classes of diagrams that contribute.
The net result is a vanishing effect.
Thus, even if the rho-mass splitting
will be better known one day,
it will never be a great source of CIB.

Another CIB contribution to the nuclear
force is irreducible pion-photon ($\pi\gamma$) exchange.
Traditionally, it was believed that this contribution would take care
of the remaining 20\% of $\Delta a_{CIB}$~\cite{EM83,Ban75,Che75}.
However, a recently derived $\pi\gamma$ potential based upon chiral
perturbation theory~\cite{Kol98} {\it decreases} $\Delta a_{CIB}$
by about 0.5 fm, making the discrepancy even larger.

{\it Thus, it is a matter of fact that about 25\%
of the charge-dependence of the singlet scattering length is
not explained---at this time.}

\section{The $\pi NN$ coupling constant}

For the nuclear force,
the pion is the most important meson.
Therefore, it is crucial to have an accurate understanding of the
coupling of the pion to the nucleon.
In the 1990's, we have seen a controversial discussion
about the precise value for the $\pi NN$ coupling constant.
We will first briefly review the events and then discuss
in which way the NN data impose constraints on this
important coupling constant.

From 1973 to 1987, there was a consensus that the $\pi NN$
coupling constant is $g^2_\pi/4\pi=14.3\pm 0.2$
(equivalent to
$f^2_{\pi}=0.079 \pm 0.001$\footnote{Using $\pi NN$ 
Lagrangians as defined in the authoritative review
\cite{Dum83}, the relevant relationships between the pseudoscalar pion coupling
constant, $g_\pi$, and the pseudovector one, $f_\pi$, are
\begin{equation}
\frac{g^2_{\pi^0 pp}}{4\pi} = 
      \left( \frac{2M_p}{m_{\pi^\pm}} \right)^2 f^2_{\pi^0 pp}
      = 180.773 f^2_{\pi^0 pp}
\end{equation}
and
\begin{equation}
\frac{g^2_{\pi^\pm}}{4\pi} = 
      \left( \frac{M_p+M_n}{m_{\pi^\pm}} \right)^2 f^2_{\pi^\pm}
       = 181.022 f^2_{\pi^\pm} \; .
\end{equation}
with $M_p=938.272$ MeV the proton mass,
$M_n=939.566$ MeV the neutron mass, and
$m_{\pi^\pm}=139.570$ MeV the mass of the charged pion.}).
This value was obtained by Bugg {\it et al.}~\cite{BCC73}
from the analysis of $\pi^\pm p$ data in 1973, and
confirmed by
Koch and Pietarinen~\cite{KP80} in 1980.
Around that same time,
the neutral-pion coupling constant
was determined by Kroll~\cite{Kro81}
from the analysis of $pp$ data by means of forward dispersion
relations; he obtained
$g^2_{\pi^0}/4\pi = 14.52 \pm 0.40$ 
(equivalent to
$f^2_{\pi^0} = 0.080 \pm 0.002$).

The picture changed in 1987, when the Nijmegen group~\cite{Ber87}
determined the neutral-pion coupling constant in a partial-wave analysis
of $pp$ data and obtained 
$g^2_{\pi^0}/4\pi = 13.1 \pm 0.1$.
Including also the magnetic moment interaction between protons in the analysis,
the value shifted to $13.55 \pm 0.13$ in 1990~\cite{Ber90}.
Triggered by these events, Arndt {\it et al.}~\cite{Arn90} reanalysed
the $\pi^\pm p$ data to determine the charged-pion coupling constant
and obtained 
$g^2_{\pi^\pm}/4\pi = 13.31 \pm 0.27$.
In subsequent work, the Nijmegen group also analysed $np$, $\bar{p}p$,
and $\pi N$ data~\cite{Tim95}. The status of their work as of 1993 is summarized in
Ref.~\cite{STS93} where they claim that the most accurate values 
are obtained in their combined $pp$ and $np$ analysis yielding
$g^2_{\pi^0}/4\pi = 13.47 \pm 0.11$ 
(equivalent to 
$f^2_{\pi^0} = 0.0745 \pm 0.0006$)
and
$g^2_{\pi^\pm}/4\pi = 13.54 \pm 0.05$ 
(equivalent to
$f^2_{\pi^\pm}=0.0748 \pm 0.0003$).
The latest analysis of all $\pi^\pm p$ data below 2.1 GeV conducted by
the VPI group
using fixed-$t$ and forward dispersion relation constraints
has generated
$g^2_{\pi^\pm}/4\pi=13.75\pm 0.15$~\cite{AWP94}.
The VPI NN analysis extracted
$g^2_{\pi^0}/4\pi \approx 13.3$ and 
$g^2_{\pi^\pm}/4\pi \approx 13.9$ as well as
the charge-independent value
$g^2_{\pi}/4\pi \approx 13.7$  
\cite{ASW94,ASW95}.

Also Bugg and coworkers
have performed new determinations of the $\pi NN$ coupling
constant. Based upon precise $\pi^\pm p$ data in the 100--310 MeV range
and applying fixed-$t$ dispersion relations, they obtained the value
$g^2_{\pi^\pm}/4\pi=13.96\pm 0.25$
(equivalent to 
$f^2_{\pi^\pm} = 0.0771 \pm 0.0014$)~\cite{MB93}.
From the analysis of NN elastic data between
210 and 800 MeV,
Bugg and Machleidt~\cite{BM95} have deduced
$g^2_{\pi^\pm}/4\pi = 13.69\pm 0.39$ and
$g^2_{\pi^0}/4\pi = 13.94\pm 0.24$.

Thus, it may appear that recent determinations show a consistent
trend towards a lower value for $g_\pi$ with no indication for
substantial charge dependence.

However, this is not true and for a comprehensive
overview of recent determinations of the $\pi NN$
coupling constant, see reference~\cite{Upp99}. In particular, there is
one determination that does not follow the above trend.
Using a modified Chew extrapolation procedure, 
the Uppsala Neutron Research Group has deduced 
the charged-pion coupling constant from high
precision $np$ charge-exchange data at 162 MeV~\cite{Eri95}.
Their latest result is
$g^2_{\pi^\pm}/4\pi = 14.52\pm 0.26$~\cite{Rah98}.
We note that the method used by the Uppsala Group 
is controversial~\cite{RKS98,SRT98}.

Since the pion plays a crucial role in the creation of
the nuclear force, many NN observables are sensitive
to the $\pi NN$ coupling constant, $g_\pi$.
We will discuss here the most prominent cases and their
implications for an accurate value of $g_\pi$.

We will focus on
the deuteron, NN analyzing powers $A_y$, and the singlet scattering
length. Other NN observables with sensitivity to
$g_\pi$ are spin transfer coefficients. Concerning
the latter and their implications for $g_\pi$, we refer
the interested reader to references~\cite{BM95,Wi99}.

\begin{table}
\caption{Important coupling constants and the predictions for the deuteron
and some $pp$ phase shifts for five models discussed in the text.}
\begin{indented}
\item[]\begin{tabular}{lllllll}
\br
   &~A~&~B~&~C~&~D~&~E~&Empirical\\
\br
\multicolumn{7}{c}{\bf Important coupling constants}\\
~$g^2_{\pi^0}/4\pi$~&~13.6~&~13.6~&~14.0~&~14.4~&~13.6~&~\\
~$g^2_{\pi^\pm}/4\pi$~&~13.6~&~13.6~&~14.0~&~14.4~&~14.4~&~\\
~$\kappa_\rho$~&~6.1~&~3.7~&~6.1~&~6.1~&~6.1~&~\\
\mr
\multicolumn{7}{c} {\bf The deuteron}\\
~$Q$ 
(fm$^2$)&~0.270~&~0.278~&~0.276~&~0.282~&~0.278~&~0.276(2)$^a$~\\
~$\eta$&~0.0255~&~0.0261~&~0.0262~&~0.0268~&~0.0264
&~0.0256(4)$^b$~\\
$A_S$ (fm$^{-1/2}$)&0.8845&0.8842&0.8845&0.8845&0.8847&0.8845(8)$^c$\\
~$P_D$ (\%)&~4.83~&~5.60~&~5.11~&~5.38~&~5.20~&~--\\
\mr
\multicolumn{7}{c}{{\bf \mbox{\boldmath $^3P_0$ $pp$} phase shifts} (deg)}\\
10 MeV & 3.726 & 4.050 & 3.881 & 4.039 & 3.726  & 3.729(17)$^d$\\  
25 MeV & 8.588 & 9.774 & 8.981 & 9.384 & 8.588  & 8.575(53)$^d$\\
50 MeV &11.564&14.070&12.158& 12.763&11.564 & 11.47(9)$^d$ \\
\br
\end{tabular}
\item[]
$^a$ Corrected for meson-exchange currents 
and relativity.
\item[]
$^b$ Reference~\cite{RK90}.
\item[]
$^c$ Reference~\cite{ER83}.
\item[]
$^d$ Nijmegen $pp$ multi-energy phase shift analysis~\cite{Sto93}.
\end{indented}
\label{tab_deu}
\end{table}

\subsection{The deuteron}

The crucial deuteron observables to consider are the quadrupole moment,
$Q$, and the asymptotic D/S state ratio, $\eta$.
The sensitivity of both quantities to $g_\pi$ is demonstrated in 
table~\ref{tab_deu}. 
The calculations are based upon the CD-Bonn potential~\cite{MSS96,Mac00}) 
which belongs to the new generation
of high-precision NN potentials that fit the NN data below
350 MeV with a `perfect' $\chi^2$/datum of about one.
The numbers in table~\ref{tab_deu} are an update of earlier calculations
of this kind~\cite{MS91,ML93} in which older NN potentials
were applied. However, there are no substantial differences in the results
as compared to the earlier investigations.

For meaningful predictions, it is important that all deuteron models
considered are realistic. This requires that besides the deuteron
binding energy (that is accurately reproduced by all models of
table~\ref{tab_deu}) also other empirically well-known quantities 
are correctly predicted, like
the deuteron radius, $r_d$,  and the triplet effective range
parameters, $a_t$ and $r_t$. As it turns out,
the latter quantities are closely related to the asymptotic S-state
of the deuteron, $A_S$, which itself is not an observable.
However, it has been shown~\cite{ER83} that for realistic values
of $r_d$, $a_t$, and $r_t$, the asymptotic S-state of the deuteron
comes out to be in the range $A_S=0.8845\pm 0.0008$ fm$^{-1/2}$.
Thus, $A_S$ plays the role of 
an important control number that tells us if a deuteron
model is realistic or not. As can be seen from table~\ref{tab_deu},
all our models pass the test.

\begin{figure}

\vspace*{0.5cm}
\hspace*{2.5cm}
\epsfig{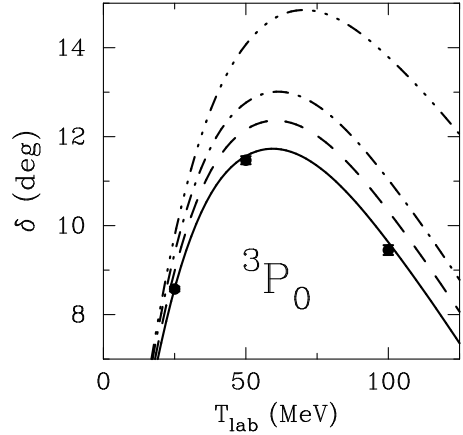}

\vspace*{0.0cm}
\caption{$^3P_0$ phase shifts of proton-proton scattering
as predicted by Model A and E ($g^2_{\pi^0}/4\pi=13.6$, 
solid line), 
B ($\kappa_\rho=3.7$, 
dash-3dot), C ($g^2_{\pi^0}/4\pi=14.0$, 
dashed), and D ($g^2_{\pi^0}/4\pi=14.4$, 
dash-dot)~\footnote{Note that in $pp$ scattering 
only the $\pi^0$ is exchanged.}.
The solid dots represent the Nijmegen $pp$ multi-energy phase shift
analysis~\cite{Sto93}.}
\label{fig_3p0}
\end{figure}

Model A of table~\ref{tab_deu} uses the currently fashionable value
for the $\pi NN$ coupling constant
$g^2_\pi/4\pi = 13.6$ which clearly underpredicts $Q$
while $\eta$ is predicted satisfactorily. 
One could now try to fix the problem with $Q$ 
by using a weaker
$\rho$-meson tensor-coupling to the nucleon, $f_\rho$.
It is customary to state the strength of this coupling in terms of the
tensor-to-vector ratio of the $\rho$ coupling constants,
$\kappa_\rho\equiv f_\rho/g_\rho$. Model A uses the `large' value
$\kappa_\rho=6.1$ recommended by Hoehler and Pietarinen~\cite{HP75}. 
Alternatively, one may try the value implied by the vector-meson
dominance model for the electromagnetic form factor of the 
nucleon~\cite{Sak69}
which is $\kappa_\rho=3.7$. This is done in our Model B which shows
the desired improvement of $Q$. However, a realistic model for the
NN interaction must not only describe the deuteron but also
NN scattering. As discussed in detail in reference~\cite{BM94},
the small $\kappa_\rho$ cannot reproduce the $\epsilon_1$ mixing
parameter correctly and, in addition, there are serious problems with
the $^3P_J$ phase shifts, particularly, the $^3P_0$ (cf.\ lower
part of table~\ref{tab_deu} and figure~\ref{fig_3p0}). 
Therefore, Model B is unrealistic and must be discarded.

The only parameters left to improve $Q$ are $g_\pi$ and the 
$\pi NN$ vertex form-factor,
$F_{\pi NN}$ 
(cf.\ equation~\ref{eq_ope}, above).
As for the $\rho$ meson, 
$F_{\pi NN}$ 
is heavily constrained by NN phase parameters, particularly,
$\epsilon_1$. The accurate reproduction of $\epsilon_1$
as determined in the Nijmegen $np$ multi-energy phase shift 
analysis~\cite{Sto93} essentially leaves no room for variations
of 
$F_{\pi NN}$ 
once the $\rho$ meson parameters are fixed.

Thus, we are finally left with only one parameter to fix the
$Q$ problem, namely $g_\pi$. As it turns out, for relatively
small changes of 
$g^2_\pi/4\pi$
there is a linear relationship, as demonstrated in table~\ref{tab_deu}
by the predictions of Model A, C and D which use
$g^2_\pi/4\pi=13.6$, 14.0, and 14.4, respectively.
Consistent with earlier studies~\cite{MS91,ML93},
one finds that 
$g^2_\pi/4\pi\geq14.0$
is needed to correctly reproduce $Q$.

However, a pion coupling with
$g^2_\pi/4\pi\geq14.0$
creates problems for the $^3P_0$ phase shifts which are
predicted too large at low energy (cf.\ lower part of table~\ref{tab_deu}
and figure~\ref{fig_3p0}).
Now, a one-boson-exchange (OBE) model for the NN interaction
includes several parameters (about one dozen in total).
One may therefore try to improve the
$^3P_0$ by readjusting some of the other model parameters.
The vector mesons ($\rho$ and $\omega$) have a strong impact on the
$^3P_0$ (and the other $P$ waves).
However, due to their heavy masses, they are more effective at high
energies than at low ones. Therefore, $\rho$ and $\omega$ 
may produce large changes of the $^3P_0$ phase shifts
in the range 200-300 MeV, with
little improvement at low energies. The bottom line is that
in spite of the large number of parameters in the model,
there is no way to fix the $^3P_0$ phase shift at low energies. In this
particular partial wave, 
the pion coupling constant is the only effective parameter, 
at energies below 100 MeV.
The $pp$ phase
shifts of the Nijmegen analysis~\cite{Sto93} as well as 
the $pp$ phases produced by the VPI group~\cite{SAID} require
$g^2_\pi/4\pi\leq13.6$.

Notice that this finding is in clear contradiction to our conclusion
from the deuteron $Q$.

There appears to be a way to resolve this problem.
One may assume that the neutral pion, $\pi^0$, couples to the
nucleon with a slightly different strength than the charged
pions, $\pi^\pm$. This assumption of a charge-splitting of
the $\pi NN$ coupling constant is made in our Model E where
we use
$g^2_{\pi^0}/4\pi=13.6$ and
$g^2_{\pi^\pm}/4\pi=14.4$.
This combination reproduces the $pp$ $^3P_0$ phase shifts at low
energy well
and creates a sufficiently large deuteron $Q$.

\subsection{Analyzing powers}

\begin{figure}

\vspace*{0.5cm}
\hspace*{2.5cm}
\epsfig{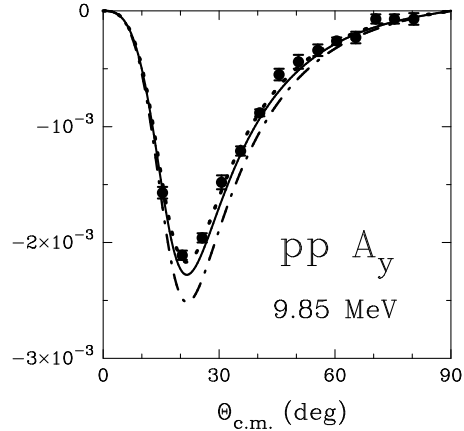}

\caption{The proton-proton analyzing power $A_y$
at 9.85 MeV.
The theoretical curves are calculated with
$g^2_{\pi^0}/4\pi=13.2$ (dotted), 
13.6 (solid, Model~A), and 14.4 (dash-dot, Model~D)
and fit the data with a $\chi^2$/datum of
0.98, 2.02, and 9.05, respectively.
The solid dots represent the data taken at
Wisconsin~\cite{Bar82}.}
\label{fig_aypp}
\end{figure}

In our above considerations, 
some $pp$ phase shifts played
an important role. 
In principle, phase shifts are
nothing else but an alternative representation of data.
Thus, one may as well use the data directly. 
Since the days of Gammel and Thaler~\cite{GT57}, it is well-known
that the triplet $P$-wave phase shifts
are fixed essentially by the NN
analyzing powers, $A_y$. Therefore, we will now take a look
at $A_y$ data and compare them directly with model predictions.

\begin{table}[t]
\caption{$\chi^2$/datum for the fit of the world 
$pp$ $A_y$ data below 350 MeV
(subdivided into three energy ranges) using different values
of the $\pi NN$ coupling constant.}
\begin{indented}
\item[]\begin{tabular}{lcccc}
\br
  & \multicolumn{4}{c}{\bf Coupling constant \mbox{\boldmath $g^2_{\pi^0}/4\pi$}}\\
Energy range (\# of data) & 13.2 & 13.6 & 14.0 & 14.4 \\
   &   & A & C & D \\
\mr
0--17 MeV (45 data) & 0.84 & 1.43 & 2.71 & 4.66 \\
17--125 MeV (148 data) & 1.05 & 1.06 & 1.54 & 2.45 \\
125--350 MeV (624 data) & 1.24 & 1.22 & 1.26 & 1.34 \\
\br
\end{tabular}
\end{indented}
\label{tab_aypp}
\end{table}

In figure~\ref{fig_aypp}, we show high-precision $pp$ $A_y$ data at 9.85 MeV
from Wisconsin~\cite{Bar82}.
The theoretical curves shown are obtained with
$g^2_{\pi^0}/4\pi=13.2$ (dotted), 13.6 (solid), and 14.4 (dash-dot)
and fit the data with a $\chi^2$/datum of
0.98, 2.02, and 9.05, respectively.
Clearly, a small coupling constant around 13.2 is favored.
Since a single data set is not a firm basis, we have looked into
all $pp$ $A_y$ data in the energy range 0--350 MeV.
Our results are presented in table~\ref{tab_aypp} 
where we give the $\chi^2$/datum
for the fit of the world $pp$ $A_y$ data below 350 MeV
(subdivided into three energy ranges) for various choices
of the neutral $\pi NN$ coupling constant.
It is seen that the $pp$ $A_y$ data at low energy, particularly
in the energy range 0--17 MeV, are very sensitive to the
$\pi NN$ coupling constant.
A value $g^2_{\pi^0}/4\pi \leq 13.6$ is clearly preferred, consistent
with what we extracted from the single data set at 9.85 MeV as well as
from the $^3P_0$ phase shifts.

\begin{figure}

\vspace*{0.5cm}
\hspace*{2.5cm}
\epsfig{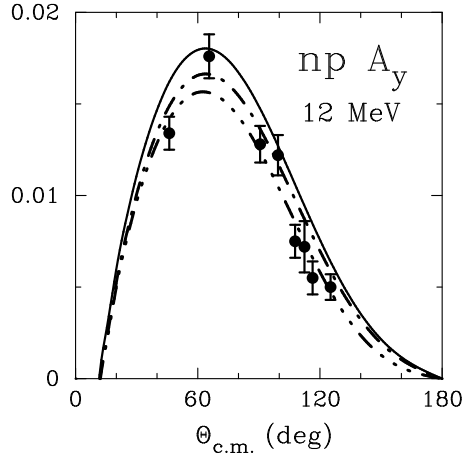}

\caption{The neutron-proton analyzing power $A_y$
at 12 MeV.
The theoretical curves are calculated with
$g^2_{\pi^0}/4\pi
=g^2_{\pi^\pm}/4\pi
=13.6$ (solid line, Model A), 
$g^2_{\pi^0}/4\pi
=g^2_{\pi^\pm}/4\pi
=14.4$ (dash-dot, Model D), and the charge-splitting
$g^2_{\pi^0}/4\pi=13.6,\;
g^2_{\pi^\pm}/4\pi=14.4$ (dash-3dot, Model E). 
The solid dots represent the data taken at
TUNL~\cite{Wei92}.}
\label{fig_aynp}
\end{figure}

Next, we look into the $np$ $A_y$ data.
A single sample is shown in figure~\ref{fig_aynp},
the $np$ $A_y$ data at 12 MeV from TUNL~\cite{Wei92}.
Predictions are shown for Model A (solid line), D (dash-dot),
and E (dash-triple-dot). The charge-splitting Model E
fits the data best with a $\chi^2$/datum of 1.00 
(cf.\ table~\ref{tab_aynp}).
We have also considered the entire $np$ $A_y$ data measured
by the TUNL group~\cite{Wei92} in the energy range 7.6--18.5 MeV
(31 data) as well as 
the world $np$ $A_y$ data
in the energy ranges 0--17 MeV (120 data).
It is seen that there is some sensitivity to the $\pi NN$
coupling constant in this energy range, while there is little
sensitivity at energies above 17 MeV (cf.\ table~\ref{tab_aynp}).

\begin{table}[b]
\caption{$\chi^2$/datum for the fit of various sets of $np$ $A_y$ data
using different values
for the $\pi NN$ coupling constants.}
\footnotesize
\begin{tabular}{lccccc}
\br
  & \multicolumn{5}{c}{\bf Coupling constants \mbox{\boldmath 
$g^2_{\pi^0}/4\pi; \; g^2_{\pi^\pm}/4\pi$}}\\
Energy, data set (\# of data) &~13.6; 13.6~&~14.0; 14.0~&~14.4; 14.4
 &~13.6; 14.0~&~13.6; 14.4~\\
   &~A~&~C~&~D~&   &~E~\\
\br
12 MeV \cite{Wei92} (9 data)
 & 2.81 & 2.27 & 1.79 & 1.53 & 1.00 \\
7.6--18.5 MeV \cite{Wei92} (31 data)
 & 1.89 & 1.56 & 1.29 & 1.28 & 1.32 \\
0--17 MeV world data (120)
 & 1.17 & 1.03 & 0.94 & 0.99 & 1.19 \\
\mr
17--50 MeV \cite{Wil84} (85 data)
 & 1.16 & 1.12 & 1.14 & 1.18 & 1.18 \\
17--125 MeV world data (416)
 & 0.89 & 0.89 & 0.91 & 0.91 & 0.94 \\
\br
\end{tabular}
\label{tab_aynp}
\end{table}

Consistent with the trend seen in the 12 MeV data, 
the larger data sets below 17 MeV show a clear preference for
a coupling constant around 14.4 if there is no charge splitting
of $g_\pi$. 
This implies that without charge-splitting
it is impossible to obtain
an optimal fit of the $pp$ and $np$
$A_y$ data. To achieve this best fit, charge-splitting is needed,
like $g^2_{\pi^0}/4\pi=13.6$
and $g^2_{\pi^\pm}/4\pi=14.0$,
as considered in column 5 of table~\ref{tab_aynp}.
The drastic charge-splitting of Model E is not favored by the
more comprehensive $np$ $A_y$ data sets.

The balance of the analysis of the $pp$ and $np$ $A_y$ data then is: 
$g^2_{\pi^0}/4\pi \leq 13.6$
and $g^2_{\pi^\pm}/4\pi \geq 14.0$.
Notice that this
splitting is consistent with our conclusions from the deuteron.
Thus, we have now some indications for charge-splitting of $g_\pi$
from two very different observables, namely the deuteron quadrupole
moment and $np$ analyzing powers.

Therefore, it is worthwhile to look deeper into the issue
of charge-splitting of the $\pi NN$ coupling constant.
Unfortunately, there are severe problems with any substantial
charge-splitting---for two reasons.
First, theoretical work~\cite{Hwa98} on isospin symmetry breaking
of the $\pi NN$ coupling constant based upon QCD sum rules comes
up with a splitting of less than 0.5\% for $g_\pi^2$ and, thus,
cannot explain the large charge splitting indicated above.
Second, a problem occurs with the conventional
explanation of the charge-dependence of the singlet
scattering length, which we will explain now.

\subsection{Charge-dependence of the singlet scattering length and
charge-dependence of the pion coupling constant}

Here, we are going to show in detail how 
charge-splitting of the $\pi NN$ coupling constant
affects the charge-dependence of the $^1S_0$ scattering
length. It will turn out that the suggested charge-splitting of $g_\pi$
causes a disaster for our established
understanding of the charge-dependence of the singlet
scattering length. 

Our above considerations suggest charge-splitting
of $g_\pi$, like
\begin{eqnarray}
g^2_{\pi^0}/4\pi & = & 13.6 \; ,
\label{eq_cdgpi1}
\\
g^2_{\pi^\pm}/4\pi & = & 14.4 \: ,
\label{eq_cdgpi2}
\end{eqnarray}
cf.\ Model E of table~\ref{tab_deu}.
We will now discuss how this charge-splitting of $g_\pi$ affects
$\Delta a_{CIB}$ (more details can be found in the original paper
reference~\cite{MB99}).

Accidentally, this splitting is---in relative terms---about the same as the
pion-mass splitting; that is
\begin{equation}
\frac{g_{\pi^0}}{m_{\pi^0}} \approx
\frac{g_{\pi^\pm}}{m_{\pi^\pm}}\; . 
\label{eq_cdgpi3}
\end{equation}
As discussed (cf.\ equations (\ref{eq_pi1s0}) and (\ref{eq_2pi})
and text below these equations), 
for zero momentum transfer, we have roughly
for one-pion exachange
\begin{equation}
\mbox{OPE} \sim \left(\frac{g_\pi}{m_\pi}\right)^2
\end{equation}
and
for $2\pi$ exchange
\begin{equation}
\mbox{TPE} \sim \left(\frac{g_\pi}{m_\pi}\right)^4 \; ,
\end{equation}
which is not unexpected, anyhow.
On the level of this qualitative discussion, we can then predict that
any pionic charge-splitting 
satisfying equation~(\ref{eq_cdgpi3}) will create no CIB from pion exchanges.
Consequently, a charge-splitting of $g_\pi$ as given in 
equations~(\ref{eq_cdgpi1})
and (\ref{eq_cdgpi2})
will wipe out our established explanation of CIB
of the NN interaction.

In reference~\cite{MB99}, accurate numerical calculations based upon the Bonn
meson-exchange model for the NN interaction~\cite{MHE87} have been conducted.
The details of these calculations are spelled out
in reference~\cite{LM98b} where, however, no charge-splitting of $g_\pi$
was considered. Assuming the $g_\pi$ of 
equations~(\ref{eq_cdgpi1})
and (\ref{eq_cdgpi2}),
one obtains the $\Delta a_{CIB}$ 
predictions given in the last column of table~\ref{tab_cibgpi}.
It is seen that the results of an accurate calculation go even beyond
what the qualitative estimate suggested:
the conventional CIB prediction is not only reduced, it is reversed.
This is easily understood if one recalls [cf.\ equations~(\ref{eq_pi1s0})
and (\ref{eq_2pi})] that the pion mass appears
in the propagator $(m_\pi^2+{\bf k}^2)^{-1}$. Assuming an
average ${\bf k}^2\approx m^2_\pi$, the 7\% charge splitting of
$m^2_\pi$ will lead to only about a 3\% charge-dependent effect from
the propagator. Thus, if a 6\% charge-splitting of $g_\pi^2$ is used, 
this will not only override the pion-mass effect, it will reverse it.

Based upon this argument and on the numerical results, 
one can then estimate that
a charge-splitting of $g_\pi^2$ of only about 3\%
(e.~g., 
$g^2_{\pi^0}/4\pi = 13.6$ and $g^2_{\pi^\pm}/4\pi = 14.0$)
would erase all predictions of CIB in the singlet scattering length
derived from pion mass splitting.

Besides pion mass splitting, we do not know of any other essential mechanism
to explain the charge-dependence of the singlet scattering length. 
Therefore, it is unlikely
that this mechanism is annihilated by a charge-splitting of $g_\pi$.
This may be taken as an indication that there is no significant
charge splitting of the $\pi NN$ coupling constant.

\begin{table}
\caption{Predictions for $\Delta a_{CIB}$ 
in units of fm without and with the assumption of charge-dependence
of $g_\pi$.}
\begin{indented}
\item[]\begin{tabular}{lcc}
\br
       & {\bf No charge-dependence of $g_\pi$} 
       & {\bf Charge-dependent $g_\pi$:} 
\\ 
       & $g^2_{\pi^0}/4\pi = g^2_{\pi^\pm}/4\pi = 14.4$
       & $g^2_{\pi^0}/4\pi = 13.6$
\\
       &   
       & $g^2_{\pi^\pm}/4\pi = 14.4$
\\
\br
$1\pi$  & 3.24 & --1.58 \\
$2\pi$  & 0.36 & --1.94 \\
$\pi\rho,\pi\sigma,\pi\omega$ & 1.04 & --0.97 \\
\mr
Sum &  4.64 & --4.49 \\
\br
Empirical & \multicolumn{2}{c}{$5.64\pm 0.60$}\\
\br
\end{tabular}
\end{indented}
\label{tab_cibgpi}
\end{table}

\subsection{Conclusions}

Several NN observables can be identified that are very 
sensitive to the $\pi NN$ coupling constant, $g_\pi$.
They all carry the potential to determine
$g_\pi$ with high precision.

In particular, we have shown that the $pp$ $A_y$ data below
17 MeV are very sensitive to $g_\pi$ and imply a value
$g^2_\pi/4\pi \approx 13.2$.
The $np$ $A_y$ data below 17 MeV show moderate sensitivity
and the deuteron quadrupole moment shows great sensitivity to $g_\pi$;
both $np$ observables imply $g^2_\pi/4\pi\geq 14.0$.

The two different values may suggest a relatively large
charge-splitting of $g_\pi$. However, 
a charge-splitting of this kind would completely destroy our
established explanation of the charge-dependence of the
singlet scattering length. Since this is unlikely to be true,
we must discard the possibility of any substantial charge-splitting
of $g_\pi$.

The conclusion then is that we are faced with real and substantial
discrepancies between the values for $g_\pi$ based upon
different NN observables.
The reason for this can only be that there are large,
unknown systematic errors in the data and/or large
uncertainties in the theoretical methods.
Our homework for the future is to find these errors and
eliminate them.

Another way to summarize the current confused situation is to state
that, presently, any value between 13.2 and 14.4 is possible for
$g^2_\pi/4\pi$ depending on which NN observable you pick.
If we want to pin down the value more tightly, then we are faced with
three possible scenarios:
\begin{itemize}
\item
$g_\pi$ is  small,
$g^2_\pi/4\pi\leq 13.6$:\\
The deuteron $\eta$ and $pp$ scattering
at low energies are described well; 
there are moderate problems with the $np$ $A_y$ data
below 17 MeV.
{\it The most serious problem is the deuteron $Q$.}
Meson-exchange current contributions (MEC) and relativistic
corrections for $Q$ of 0.016~fm$^2$ or more would solve the problem.
Present calculations predict about 0.010~fm$^2$ or less.
A serious reinvestigation of this issue is called for.
\item
$g_\pi$ is large,
$g^2_\pi/4\pi\geq 14.0$:\\
The deuteron $Q$ is well reproduced, but $\eta$ is predicted too large
as compared to the most recent measurement by Rodning and Knutsen~\cite{RK90},
$\eta = 0.0256(4)$.
Note, however, that all earlier measurements of $\eta$ came up
with a larger value; for example, Borbely {\it et al.}~\cite{Bor89}
obtained $\eta = 0.0273(5)$. 
There are no objectively verifiable reasons
why the latter value should be less reliable than the former one.
The deuteron $\eta$ carries the potential of being the best observable to
determine $g_\pi$ (as pointed out repeatedly by Ericson~\cite{ER83,ER84}
in the 1980's); but the unsettled experimental situation
spoils it all. 
The $np$ $A_y$ data at low energy are described well.
{\it The most serious problem are the $pp$ $A_y$ data
below 100 MeV.}
\item
$g_\pi$ is `in the middle',
$13.6 \leq g^2_\pi/4\pi \leq 14.0$:\\
we have all of the above problems,
but in moderate form.
\end{itemize}
In conclusion, to arrive at an accurate value for $g_\pi$,
there is a lot of homework to do---for theory and experiment.

\section{Phase shift analysis}

In spite of the large NN database available in the 1990's,
conventional phase shift analyses are by no means perfect.
For example, the phase shift solutions obtained
by Bugg~\cite{BB92} or the VPI/GWU group~\cite{SAID}
typically have a $\chi^2/$datum of
1.3 or more, for the energy range 0--425 MeV. 
This may be due to inconsistencies in the data
as well as deficiencies in the constraints applied in the analysis.
In any case, it is a matter of fact that
within the conventional phase shifts analysis, in which the lower partial
waves are essentially unconstrained, a better fit cannot be achieved.

About two decades ago, the Nijmegen group embarked on a program
to substantially improve NN phase shift analysis.
To achieve their goal, the Nijmegen group took two 
decisive measures~\cite{Sto93}.
First, they `pruned' the database; i.e., 
they scanned very critically the world NN database 
(all data in the energy range 0-350 MeV laboratory energy 
published in a regular physics journal
between January 1955 and December 1992) and eliminated all data that
had either an improbably high $\chi^2$ 
(more than three standard deviations off) or
an improbably low $\chi^2$; 
of the 2078
world $pp$ data below 350 MeV  
1787 survived the scan, and of the 3446 $np$ data 2514 survived.
Second, they introduced
sophisticated, semi-phenomenological model assumptions 
into the analysis. Namely,
for each of the lower partial waves ($J\leq 4$) a
different energy-dependent potential is adjusted
to constrain the energy-dependent analysis.
Phase shifts are obtained using these potentials in a Schroedinger
equation. From these phase shifts
the predictions for the observables are calculated including the $\chi^2$
for the fit of the experimental data. This $\chi^2$ is then minimized
as a function of the parameters of the partial-wave potentials.
Thus, strictly speaking, the Nijmegen analysis is a {\it potential analysis};
the final phase shifts are the ones
predicted by the `optimized' partial-wave
potentials.

In the Nijmegen analysis,
each partial-wave potential consists of a short- and a long-range
part, with the separation line at $r=1.4$ fm.
The long-range potential $V_L$ ($r>1.4$ fm) 
is made up of an electromagnetic part $V_{EM}$
and a nuclear part $V_N$: 
\begin{equation}
V_L=V_{EM}+V_N 
\end{equation}
The electromagnetic interaction
can be written as
\begin{equation}
V_{EM}(pp)=V_C+V_{VP}+V_{MM}(pp)
\end{equation}
for proton-proton scattering and
\begin{equation}
V_{EM}(np)=V_{MM}(np)
\end{equation}
for neutron-proton scattering,
where $V_C$ denotes an improved Coulomb potential
(which takes into account the lowest-order relativistic corrections
to the static Coulomb potential and includes contributions
of all two-photon exchange diagrams);
$V_{VP}$ is the vacuum polarization potential,
and $V_{MM}$ the magnetic moment interaction.

The nuclear long-range potential $V_N$
consists of the local one-pion-exchange (OPE) tail $V_{1\pi}$
(the coupling constant $g_\pi$
being one of the parameters used to minimize
the $\chi^2$) multiplied by a factor $M/E$
and the tail of the heavy-boson-exchange (HBE)
contributions of the Nijmegen78 potential~\cite{NRS78} $V_{HBE}$,
enhanced by a factor of 1.8 in singlet states; i.~e.
\begin{equation}
V_N=\frac{M}{E}\times V_{1\pi}(g_\pi,m_\pi) + f(S)\times V_{HBE}
\end{equation}
with $f(S=0)=1.8$ and $f(S=1)=1.0$,
where $S$ denotes the total spin of the two-nucleon system.
The energy-dependent factor $M/E$ (with $E=\sqrt{M^2+q^2}$, 
 $\; q^2=MT_{lab}/2$) takes into account
relativity in a `minimal' way, damping the nonrelativistic
OPE potential at higher energies.

As indicated, $V_{1\pi}$  depends on the $\pi NN$ coupling constant
$g_\pi$
 and the pion mass $m_\pi$, which gives rise to charge dependence.
For $pp$ scattering, the OPE potential is
\begin{equation}
V_{1\pi}^{pp}=V_{1\pi}(g_{\pi^0},m_{\pi^0})
\end{equation}
with $m_{\pi^0}$ the mass of the neutral pion.
In $np$ scattering, we have to distinguish between $T=1$ and $T=0$:
\begin{equation}
V^{np}_{1\pi}(T)=-V_{1\pi}(g_{\pi^0},m_{\pi^0})
+(-1)^{T+1} 2 V_{1\pi}(g_{\pi^\pm},m_{\pi^\pm})
\end{equation}

The partial-wave short-range potentials ($r \leq 1.4$ fm)
are energy-dependent square-wells (see figures~2 and 3 of reference~\cite{Sto93}).
The energy-dependence of the depth of the square-well is 
parametrized in terms of up to three parameters per
partial wave.
For the states with $J\leq 4$, there are a total of 39 such parameters
(21 for $pp$ and 18 for $np$)
plus the pion-nucleon coupling constants ($g_{\pi^0}$ and $g_{\pi^\pm}$).

In the Nijmegen $np$ analysis, the $T=1$ $np$ phase shifts are calculated
from the corresponding $pp$ phase shifts
(except in $^1S_0$ where an independent analysis is conducted) 
by applying corrections due to
electromagnetic effects and charge dependence of OPE. 
Thus, the $np$ analysis determines $^1S_0(np)$ and the
$T=0$ states, only.

In the combined Nijmegen $pp$ and $np$ analysis~\cite{Sto93}, the fit for 
1787 $pp$ data and 2514 $np$ data below 350 MeV,
available in 1993,
results in the `perfect' $\chi^2/$datum = 0.99.

\begin{table}
\caption{After-1992 proton-proton data below 350 MeV.
`Error' refers to the normalization error. This table contains 1127 observables
and 32 normalizations resulting in a total of 1159 data.}
\begin{indented}
\item[]\begin{tabular}{ccccr}
\br
                   
   $T_{lab}$ (MeV)
 & \# observable
 & Error (\%)
 & Institution(s)
 & Ref.

\\
\mr                                                                          
0.300--0.407 & 14 $\sigma$ & None & M\"unster & \cite{Do97} \\
25.68 & 8 $D$ & 1.3 & Erlangen, Z\"urich, PSI & \cite{Kr94} \\
25.68 & 6 $R$ & 1.3 & Erlangen, Z\"urich, PSI & \cite{Kr94} \\
25.68 & 2 $A$ & 1.3 & Erlangen, Z\"urich, PSI & \cite{Kr94} \\
197.4 & 41 $P$ & 1.3 &  Wisconsin, IUCF & \cite{Ra98} \\
197.4 & 41 $A_{xx}$ & 2.5 &  Wisconsin, IUCF & \cite{Ra98} \\
197.4 & 41 $A_{yy}$ & 2.5 &  Wisconsin, IUCF & \cite{Ra98} \\
197.4 & 41 $A_{xz}$ & 2.5 &  Wisconsin, IUCF & \cite{Ra98} \\
197.4 & 39 $A_{zz}$ & 2.0 &  Wisconsin, IUCF & \cite{Lo00} \\
197.8 & 14 $P$ & 1.3 &  Wisconsin, IUCF & \cite{Ha97} \\
197.8 & 14 $A_{xx}$ & 2.4 &  Wisconsin, IUCF & \cite{Ha97} \\
197.8 & 14 $A_{yy}$ & 2.4 &  Wisconsin, IUCF & \cite{Ha97} \\
197.8 & 14 $A_{xz}$ & 2.4 &  Wisconsin, IUCF & \cite{Ha97} \\
197.8 & 10 $D$ & None &  IUCF & \cite{Wi99} \\
197.8 &  5 $R$ & None &  IUCF & \cite{Wi99} \\
197.8 &  5 $R'$ & None &  IUCF & \cite{Wi99} \\
197.8 &  5 $A$ & None &  IUCF & \cite{Wi99} \\
197.8 &  5 $A'$ & None &  IUCF & \cite{Wi99} \\
250.0 & 41 $P$ & 1.3 &  IUCF, Wisconsin & \cite{Pr98} \\
250.0 & 41 $A_{xx}$ & 2.5 &  IUCF, Wisconsin & \cite{Pr98} \\
250.0 & 41 $A_{yy}$ & 2.5 &  IUCF, Wisconsin & \cite{Pr98} \\
250.0 & 41 $A_{xz}$ & 2.5 &  IUCF, Wisconsin & \cite{Pr98} \\
280.0 & 41 $P$ & 1.3 &  IUCF, Wisconsin & \cite{Pr98} \\
280.0 & 41 $A_{xx}$ & 2.5 &  IUCF, Wisconsin & \cite{Pr98} \\
280.0 & 41 $A_{yy}$ & 2.5 &  IUCF, Wisconsin & \cite{Pr98} \\
280.0 & 41 $A_{xz}$ & 2.5 &  IUCF, Wisconsin & \cite{Pr98} \\
294.4 & 40 $P$ & 1.3 &  IUCF, Wisconsin & \cite{Pr98} \\
294.4 & 40 $A_{xx}$ & 2.5 &  IUCF, Wisconsin & \cite{Pr98} \\
294.4 & 40 $A_{yy}$ & 2.5 &  IUCF, Wisconsin & \cite{Pr98} \\
294.4 & 40 $A_{xz}$ & 2.5 &  IUCF, Wisconsin & \cite{Pr98} \\
310.0 & 40 $P$ & 1.3 &  IUCF, Wisconsin & \cite{Pr98} \\
310.0 & 40 $A_{xx}$ & 2.5 &  IUCF, Wisconsin & \cite{Pr98} \\
310.0 & 40 $A_{yy}$ & 2.5 &  IUCF, Wisconsin & \cite{Pr98} \\
310.0 & 40 $A_{xz}$ & 2.5 &  IUCF, Wisconsin & \cite{Pr98} \\
350.0 & 40 $P$ & 1.3 &  IUCF, Wisconsin & \cite{Pr98} \\
350.0 & 40 $A_{xx}$ & 2.5 &  IUCF, Wisconsin & \cite{Pr98} \\
350.0 & 40 $A_{yy}$ & 2.5 &  IUCF, Wisconsin & \cite{Pr98} \\
350.0 & 40 $A_{xz}$ & 2.5 &  IUCF, Wisconsin & \cite{Pr98} \\
\br
\end{tabular}
\end{indented}
\label{tab_ppdat}
\end{table}

\begin{table}
\caption{$\chi^2$/datum for the NN data below 350 MeV
applying some recent phase shift analyses (PSA) and NN potentials.}
\footnotesize
\begin{tabular}{lcccc}
\br

 & VPI/GWU         
 & Nijmegen 1993
 & Argonne $V_{18}$
 & CD-Bonn 
\\

 & PSA~\cite{SP98}
 & PSA~\cite{Sto93}
 & pot.~\cite{WSS95}
 & pot.~\cite{Mac00}
\\
\mr 
\multicolumn{5}{c}{\bf proton-proton data}\\
1992 $pp$ database (1787 data)      & 1.28 & 1.00 & 1.10 & 1.00\\
{\bf After-1992 $pp$ data (1145 data)$^a$}&{\bf 1.08} &{\bf 1.24} &{\bf 1.74} & {\bf 1.03}\\
1999 $pp$ database (2932 data)$^a$  & 1.21 & 1.09 & 1.35 & 1.01\\
\mr 
\multicolumn{5}{c}{\bf neutron-proton data}\\
1992 $np$ database (2514 data)      & 1.19 & 0.99 & 1.08 & 1.03\\
After-1992 $np$ data (544 data)$^{b}$ & 0.98$^c$ & 0.99 & 1.02 & 0.99\\
1999 $np$ database (3058 data)$^{b}$  & 1.16$^c$ & 0.99 & 1.07 & 1.02\\
\br 
\end{tabular}
$^a$Without the 14 $pp$ $\sigma$ data of Ref.~\cite{Do97}.\\
$^{b}$Without after-1992 $np$ $\sigma$ data, except \cite{BM97}.\\
$^c$Without the data of reference~\cite{Ra99}.
\label{tab_chi2}
\end{table}

\begin{table}
\caption{After-1992 neutron-proton data below 350 MeV.
`Error' refers to the normalization error.} 
\begin{indented}
\item[]\begin{tabular}{ccccr}
\br
                   
   $T_{lab}$ (MeV)
 & \# observable
 & Error (\%)
 & Institution(s)
 & Ref.

\\
\mr                                                                          
3.65--11.6 & 9 $\Delta \sigma_T$ & None & TUNL & \cite{Wi95} \\
4.98--19.7 & 6 $\Delta \sigma_L$ & None & TUNL & \cite{Ra99} \\
4.98--17.1 & 5 $\Delta \sigma_T$ & None & TUNL & \cite{Ra99} \\
14.11      & 6 $\sigma$ & 0.7  & T\"ubingen & \cite{BM97} \\
15.8       & 1 $D_t$ & None & Bonn & \cite{Cl98} \\
16.2       & 1 $\Delta \sigma_T$ & None & Prague & \cite{Br96} \\
16.2       & 1 $\Delta \sigma_L$ & None & Prague & \cite{Br97} \\
29.0       &  6 $\sigma$ &  4.0         & Louvain-la-Neuve & \cite{Be97} \\
31.5       &  6 $\sigma$ &  4.0         & Louvain-la-Neuve & \cite{Be97} \\
34.5       &  6 $\sigma$ &  4.0         & Louvain-la-Neuve & \cite{Be97} \\
37.5       &  6 $\sigma$ &  4.0         & Louvain-la-Neuve & \cite{Be97} \\
41.0       &  6 $\sigma$ &  4.0         & Louvain-la-Neuve & \cite{Be97} \\
45.0       &  6 $\sigma$ &  4.0         & Louvain-la-Neuve & \cite{Be97} \\
49.0       &  6 $\sigma$ &  4.0         & Louvain-la-Neuve & \cite{Be97} \\
53.0       &  6 $\sigma$ &  4.0         & Louvain-la-Neuve & \cite{Be97} \\
58.5       &  6 $\sigma$ &  4.0         & Louvain-la-Neuve & \cite{Be97} \\
62.7       &  6 $\sigma$ &  4.0         & Louvain-la-Neuve & \cite{Be97} \\
67.7       & 15 $\sigma$ &  Float       & Basel, PSI    & \cite{Go94} \\
67.7       &  6 $\sigma$ &  4.0         & Louvain-la-Neuve & \cite{Be97} \\
72.8       &  6 $\sigma$ &  4.0         & Louvain-la-Neuve & \cite{Be97} \\
162.0      & 54 $\sigma$ & 2.3          & Uppsala & \cite{Ra98a} \\
175.26     & 84 $P$ & 4.9$^a$ & TRIUMF & \cite{Da96} \\
199.9     & 102 $\sigma$ & 3.0 & Freiburg, PSI & \cite{Fr99} \\
203.15     & 100 $P$ & 4.7 & TRIUMF & \cite{Da96} \\
217.24     & 100 $P$ & 4.5 & TRIUMF & \cite{Da96} \\
219.8     & 104 $\sigma$ & 3.0 & Freiburg, PSI & \cite{Fr99} \\
240.2     & 107 $\sigma$ & 3.0 & Freiburg, PSI & \cite{Fr99} \\
260.0     & 8 $R_t$ & 3.0 & PSI & \cite{Ah98} \\
260.0     & 8 $A_t$ & 3.0 & PSI & \cite{Ah98} \\
260.0     & 3 $A_t$ & 3.0 & PSI & \cite{Ah98} \\
260.0     & 8 $D_t$ & 3.0 & PSI & \cite{Ah98} \\
260.0     & 3 $D_t$ & 3.0 & PSI & \cite{Ah98} \\
260.0     & 8 $P  $ & 2.0 & PSI & \cite{Ah98} \\
260.0     & 3 $P  $ & 2.0 & PSI & \cite{Ah98} \\
261.00     &  88 $P$ & 4.1 & TRIUMF & \cite{Da96} \\
261.9     & 108 $\sigma$ & 3.0 & Freiburg, PSI & \cite{Fr99} \\
280.0     & 109 $\sigma$ & 3.0 & Freiburg, PSI & \cite{Fr99} \\
300.2     & 111 $\sigma$ & 2.6 & Freiburg, PSI & \cite{Fr99} \\
312.0     &  24 $P$ & 4.0 & SATURNE& \cite{Ba93} \\
312.0     &  11 $A_{zz}$ & 4.0 & SATURNE& \cite{Ba94} \\
318.0     & 8 $R_t$ & 3.0 & PSI & \cite{Ah98} \\
318.0     & 8 $A_t$ & 3.0 & PSI & \cite{Ah98} \\
318.0     & 5 $A_t$ & 3.0 & PSI & \cite{Ah98} \\
318.0     & 8 $D_t$ & 3.0 & PSI & \cite{Ah98} \\
318.0     & 5 $D_t$ & 3.0 & PSI & \cite{Ah98} \\
318.0     & 8 $P  $ & 2.0 & PSI & \cite{Ah98} \\
318.0     & 5 $P  $ & 2.0 & PSI & \cite{Ah98} \\
320.1     & 110 $\sigma$ & 2.1 & Freiburg, PSI & \cite{Fr99} \\
340.0     & 112 $\sigma$ & 1.8 & Freiburg, PSI & \cite{Fr99} \\
\br
\end{tabular}
\item[]
$^a$This data set is floated in the $\chi^2$ calculations of 
table~\ref{tab_chi2} because all current phase shift analyses and
$np$ potentials predict a norm that is about 4 standard deviations off
the experimental normalization error of 4.9\%. 
\end{indented}
\label{tab_npdat}
\end{table}

\begin{table}
\caption{$\chi^2$/datum for various sets of neutron-proton 
differential cross section data below 350 MeV
applying some recent phase shift analyses (PSA) and $np$ potentials.}
\footnotesize
\begin{tabular}{lcccc}
\br
                   
 & VPI/GWU 
 & Nijmegen 1993
 & Argonne $V_{18}$
 & CD-Bonn 
\\

 & PSA~\cite{SP98}
 & PSA~\cite{Sto93}
 & potential~\cite{WSS95}
 & potential~\cite{Mac00}
\\
\mr
Bonner {\it et al.}~\cite{Bo78}, 652 data & 1.18 & 1.08 & 1.20 & 1.10 \\
Freiburg/PSI~\cite{Fr99}, 871 data  & 7.66 & 8.62 & 8.58 & 8.14 \\
Uppsala~\cite{Ro92,Ra98a}, 109 data&3.40 & 6.45 & 5.20 & 6.41 \\
Louvain-la-Neuve~\cite{Be97}, 84 data & 3.22 & 3.15 & 3.12 & 3.17 \\
\br 
\end{tabular}
\label{tab_npsig}
\end{table}

\section{Recent additions to the NN database} 

The world NN data (below 350 MeV) published before December 1992
have been listed and analyzed carefully by the Nijmegen
group in their papers about the Nijmegen phase shift 
analysis~\cite{Ber90,Sto93}.
Therefore, we will focus here on data published after 1992.

\subsection{Proton-proton data}
In the past decade, there has been a major breakthrough in 
the development of experimental methods for conducting
hadron-hadron scattering experiments.
In particular, the method of internal polarized gas targets
applied in stored, cooled beams is now working perfectly in several hadron
facilities, e.~g., IUCF (Indiana, USA) and COSY (J\"ulich, Germany).
Using this new technology, IUCF has produced a large number
of $pp$ spin correlation parameters of very high precision.
In table~\ref{tab_ppdat}, we list the new IUCF data
together with other $pp$ data below 350 MeV
published between January 1993 and December 1999. 
The total number of after-1992 $pp$ data is 1159, which should be
compared to the number of $pp$ data in the (Nijmegen) 1992 base, namely, 1787. 
Thus, the $pp$ database has increased by about
2/3 since 1992. The importance of the new $pp$ data is further enhanced 
by the fact that they are of much higher quality than the old ones.

The $\chi^2$/datum produced by some recent phase shift analyses (PSA)
 and NN potentials in regard to the old and new databases are given 
in table~\ref{tab_chi2}.
In this table, the `1992 database' is the Nijmegen
database~\cite{Ber90,Sto93} and the `1999 database' is the sum of the
1992 base and the after-1992 data.

What stands out in table~\ref{tab_chi2} are the rather large values
for the $\chi^2$/datum generated by the Nijmegen analysis and the
Argonne potential for the the after-1992 $pp$ data, which are
essentially the new IUCF data. This fact is a clear indication that
these new data provide a very critical test/constraint for any NN
model. It further indicates that fitting the pre-1993 $pp$
data does not nessarily imply a good fit of those IUCF data. 
On the other hand, fitting the new IUCF data does imply a good fit of the
pre-1993 data. The conclusion from these two facts is that the new
IUCF data provide information that was not contained in the old database.
Or, in other words, the pre-1993 data were insufficient and still left
too much lattitude for pinning down NN models.
One thing in particular that we noticed is that the $^3P_1$ phase
shifts above 100 MeV have to be lower than the values given in
the Nijmegen analysis.

\subsection{Neutron-proton data}
Neutron-proton data published between January 1993 and December 1999
are listed in 
table~\ref{tab_npdat}. 
Particular attention deserve the TUNL data on $\Delta \sigma_L$
and $\Delta \sigma_T$ between 5 and 20 MeV~\cite{Ra99} which made it
possible to pin down the $\epsilon_1$ mixing parameter 
with unprecedented precision.

Table~\ref{tab_npdat} 
includes several new measurements
of $np$ differential cross sections ($\sigma$),
with the largest sets produced by the Freiburg group~\cite{Fr99}
(871 data between 200 and 350 MeV), the Uppsala group~\cite{Ro92,Ra98a}
(92 and 162 MeV, 109 data), and at Louvain-la-Neuve~\cite{Be97}
(84 data between 29 and 73 MeV). In 
table~\ref{tab_npsig} 
we show
the $\chi^2$/datum for the reproduction of these data sets by some
recent PSA and NN potentials. One observes that none of the PSA and
potentials can reproduce these data accurately.
For comparison, 
table~\ref{tab_npsig} 
includes the $\chi^2$/datum for
the $np$ $\sigma$ data by Bonner and coworkers~\cite{Bo78}
(652 data between 162 and 344 MeV, published in 1978) 
which are well reproduced.
The large differences in the $\chi^2$
implies that there are inconsistencies in the data.
Since the PSA and potentials of table~\ref{tab_npsig} were
fitted to a database that includes the Bonner data but excludes
the Freiburg data, 
one might suspect that this could be the explanation of
the good $\chi^2$ for the Bonner data and the bad one for Freiburg.
Arndt has investigated this question~\cite{Arn99} and found that
this is not entirely true.
When he excludes the Bonner data from the VPI/GWU analysis and uses
the Freiburg data instead, the latter can be reproduced with
a $\chi^2$/datum $= 2.64$ (the Bonner data produce
$\chi^2$/datum $= 1.84$ for this fit). 
Thus, the Freiburg data cannot be reproduced with the same accuracy
as the Bonner data, even if one restricts the 
$np$ $\sigma$ data exclusively to Freiburg.
This may be seen as an indication that there are inconsistencies 
within the Freiburg data. Certainly, 
the Bonner data and the Freiburg data are inconsistent with each other.
Similar problems are observed
with the Uppsala data, which are included in the VPI/GWU analysis 
and excluded from the Nijmegen PSA.
The problems with the $np$ differential cross sections deserve
further systematic investigation.

\section{The new high-precision NN potentials}

In the 1990's, a focus has been on the 
quantitative aspect of NN potentials.
Even the best NN models of the 1980's~\cite{MHE87,Lac80}
fit the NN data typically with a $\chi^2$/datum $\approx 2$ or more.
This is still substantially above the perfect
$\chi^2$/datum $\approx 1$. 
To put microscopic nuclear structure theory to a reliable test,
one needs a perfect NN potential such that discrepancies in the
predictions cannot be blamed on a bad fit of the NN data.

Based upon the Nijmegen analysis and the (pruned)
Nijmegen database, new charge-dependent NN potentials were
constructed in the early/mid 1990's.
The groups involved and the names of their new creations are,
in chronological order:
\begin{itemize}
\item
Nijmegen group~\cite{Sto94}: Nijm-I, Nijm-II, and Reid93 potentials.
\item
Argonne group~\cite{WSS95}: $V_{18}$ potential.
\item
Bonn group~\cite{MSS96,Mac00}: CD-Bonn potential.
\end{itemize}
All these potentials have in common that they use about 45 parameters and
fit the (pruned) 1992 Nijmegen data
base with a $\chi^2$/datum $\approx 1$.
However, as discussed in the previous section,
since 1992 the $pp$ database has substantially expanded
and for the current database the $\chi^2$/datum produced by some
potentials is not so perfect anymore (cf.\ table~\ref{tab_chi2}).

\subsection{Theoretical aspects}
Concerning the theoretical basis of these potential, 
one could say that they are all---more or less---constructed
`in the spirit of meson theory' (e.g., all potentials include
the one-pion-exchange contribution). However, there are
considerable differences in the details leading to considerable
off-shell differences among the potentials.

To explain these details and differences in a systematic way,
let us first sketch the general scheme for the derivation
of a meson-theoretic potential.

One starts from field-theoretic 
Lagrangians for meson-nucleon coupling, which are essentially
fixed by symmetries. Typical examples for such Langrangians are:
\begin{eqnarray}
{\cal L}_{ps}&=& -g_{ps}\bar{\psi}
i\gamma^{5}\psi\varphi^{(ps)}
\\
{\cal L}_{s}&=& -g_{s}\bar{\psi}\psi\varphi^{(s)}
\\
{\cal L}_{v}&=&-g_{v}\bar{\psi}\gamma^{\mu}\psi\varphi^{(v)}_{\mu}
-\frac{f_{v}}{4M} \bar{\psi}\sigma^{\mu\nu}\psi(\partial_{\mu}
\varphi_{\nu}^{(v)}
-\partial_{\nu}\varphi_{\mu}^{(v)})
\end{eqnarray}
where
$ps$, $s$, and $v$ denote pseudoscalar, scalar, and vector
couplings/fields, respectively.

The lowest order contributions to the nuclear force from the above
Lagrangians are the second-order Feynman diagrams
which, in the center-of-mass frame of the two interacting nucleons, produce
the amplitude:
\begin{equation}
{\cal A}_{\alpha}(q',q) =
\frac{\bar{u}_1({\bf q'})\Gamma_1^{(\alpha)} u_1({\bf q}) P_\alpha
\bar{u}_2(-{\bf q'})\Gamma_2^{(\alpha)} u_2(-{\bf q})}
{(q'-q)^2-m_\alpha^2} \; ,
\label{eq_amp}
\end{equation}
where $\Gamma_i^{(\alpha)}$ ($i=1,2$) are vertices
derived from the above Lagrangians, $u_i$ are Dirac spinors representing
the nucleons, and $q$ and $q'$ are the nucleon relative momenta
in the initial and final states, respectively; $P_\alpha$
divided by the denominator is  the
meson propagator.

The simplest meson-exchange model for the nuclear force is
the one-boson-exchange (OBE) potential~\cite{Mac89} which sums over
several second-order diagrams, each representing 
the single exchange of a different boson, $\alpha$: 
\begin{equation}
V({\bf q'},{\bf q}) =
\sqrt{\frac{M}{E'}}
\sqrt{\frac{M}{E}}
\sum_\alpha i{\cal A}_{\alpha}({\bf q'},{\bf q}) 
F_{\alpha}^2({\bf q'},{\bf q}) \; .
\label{eq_obep}
\end{equation}
As customary, we include form factors,
$F_{\alpha}({\bf q'},{\bf q})$, applied to the
meson-nucleon vertices, and a square-root factor
$M/
\sqrt{E'E}$
(with $E=\sqrt{M^2+{\bf q}^2}$
and $E'=\sqrt{M^2+{\bf q'}^2}$; $M$ is the nucleon mass).
The form factors regularize the amplitudes for large momenta
(short distances) and account for the extended structure of
nucleons in a phenomenological way.
The square root factors make it possible to cast the unitarizing,
relativistic, three-dimensional Blankenbecler-Sugar equation
for the scattering amplitude 
(a reduced version of the four-dimensional Bethe-Salpeter equation) 
into a form
which is identical to the (nonrelativistic) Lippmann-Schwinger equation
(see reference~\cite{Mac89} for details). 
Thus, equation~(\ref{eq_obep}) defines a relativistic
potential which can be consistently applied in conventional, nonrelativistic
nuclear structure.

\begin{figure}

\vspace*{-2.0cm}
\hspace*{0.75cm}
\epsfig{file=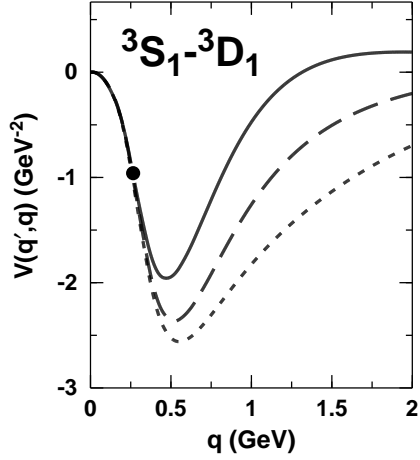,height=12.0cm}

\vspace*{-3.0cm}
\caption{Half off-shell $^3S_1$--$^3D_1$ amplitude
for the relativistic CD-Bonn potential (solid line), 
equation~(\ref{eq_obep}).
The dashed curve is obtained when the local
approximation, equation~(\ref{eq_opeq}), is used for OPE, and the dotted curve
results when this approximation is also used for one-$\rho$
exchange. $q'=265$ MeV/c.}
\label{fig_pot3sd1}
\end{figure}

Clearly, the Feynman amplitudes, equation~(\ref{eq_amp}), 
are in general nonlocal
expressions; i.~e., Fourier transforming them into 
configuration space
will yield functions of $r$ and $r'$, the relative
distances between the two in- and out-going nucleons,
respectively.
The square root factors create additional nonlocality.

While 
nonlocality appears quite plausible 
for heavy vector-meson exchange (corresponding to short distances),
we have to stress here
that even the one-pion-exchange (OPE) Feynman 
amplitude is nonlocal.
This is important because the pion creates the dominant part of the 
nuclear tensor force
which plays a crucial role in nuclear structure.

Applying $\Gamma^{(\pi)}=g_\pi \gamma_5$ in equation~(\ref{eq_amp}),
yields the Feynman amplitude for one-pion exchange, 
\begin{eqnarray}
i{\cal A}_\pi ({\bf q'}, {\bf q})  = & -\frac{g^2_\pi}{4M^2}
\frac{(E'+M)(E+M)}
{({\bf q'}-{\bf q})^2+m_\pi^2}
\left(
 \frac{\bi{\sigma_{1} \cdot } {\bf q'}}{E'+M}
-
 \frac{\bi{\sigma_{1} \cdot } {\bf q}}{E+M}
\right)
\nonumber \\
 & 
\times
\left(
 \frac{\bi{\sigma_{2} \cdot } {\bf q'}}{E'+M}
-
 \frac{\bi{\sigma_{2} \cdot } {\bf q}}{E+M}
\right) \; ,
\label{eq_operel}
\end{eqnarray}
where $m_\pi$ denotes the pion mass and
isospin factors are suppressed.
This is the original and correct result for OPE.

If one now introduces the drastic approximation,
\begin{equation}
E'\approx E \approx M \; ,
\end{equation}
then one obtains the momentum space representation of the {\it local} OPE,
\begin{equation}
V_{1\pi}^{(loc)}({\bf k})  =  -\frac{g_{\pi}^{2}}{4M^{2}}
 \frac{{\mbox {\boldmath $(\sigma_{1} \cdot $}} {\bf k)}
{\mbox{\boldmath $(\sigma_{2} \cdot $}} {\bf k)}}
 {{\bf k}^{2}+m_{\pi}^{2}}
\label{eq_opeq}
\end{equation}
with ${\bf k} = {\bf q'} - {\bf q}$.
Notice that on-shell, i.~e., for $|{\bf q'}|=|{\bf q}|$,
$V^{(loc)}_{1\pi}$ equals $i{\cal A}_\pi$.
Thus, the nonlocality affects the OPE potential
off-shell.

Fourier transform of equation~(\ref{eq_opeq}) 
yields the well-known local OPE potential
in $r$-space,
\begin{eqnarray}
V_{1\pi}^{(loc)}({\bf r}) = &
\frac{g^2_\pi}{12\pi} 
\left(\frac{m_\pi}{2M}\right)^2
\left[ 
\left(
\frac{e^{-m_\pi r}}{r}
-\frac{4\pi}{m_\pi^2}\delta^{(3)}({\bf r})
\right)
\mbox{\boldmath $\sigma_{1} \cdot \sigma_{2}$}
\right.
\nonumber \\
& +
\left.
\left(1+\frac{3}{m_\pi r}+\frac{3}{(m_\pi r)^2}\right)
\frac{e^{-m_\pi r}}{r}
\mbox{\boldmath $S_{12}$} 
\right] \; .
\label{eq_oper}
\end{eqnarray}
Notice, however, that this `well-established' local OPE potential
is only an approximative representation of the correct OPE Feynman
amplitude. A QED analog is the local Coulomb potential
{\it versus} the full field-theoretic one-photon-exchange
Feynman amplitude.

\begin{table}[t]
\caption{Modern high-precision NN potentials and 
their predictions for the two- and three-nucleon bound states.}
\small
\begin{tabular}{lcccccc}
\br
  & CD-Bonn &
 Nijm-I &
 Nijm-II &
 Reid93 &
 $V_{18}$ &
 {\sc Nature}\\
  & \cite{Mac00}& 
 \cite{Sto94}& 
 \cite{Sto94}& 
 \cite{Sto94}& 
 \cite{WSS95} &
             \\
\br
Character & nonlocal& mixed$^a$ & local & local & local &nonlocal \\
\mr
{\it Deuteron properties:} &&&&&& \\
Quadr.\ moment (fm$^2$) & 0.270 & 0.272 & 0.271 & 0.270 & 0.270  & 
   0.276(2)$^b$ \\
Asymptotic D/S state & 0.0256& 0.0253 & 0.0252 & 0.0251 & 0.0250 & 0.0256(4)$^c$ \\
D-state probab.\ (\%) & 4.85& 5.66&5.64 & 5.70  & 5.76     &  --    \\
\mr
{\it Triton binding (MeV):} &&&&&& \\
nonrel.\ calculation & 8.00 &7.72  & 7.62  & 7.63 & 7.62   & --     \\
relativ.\ calculation & 8.2  & --   &  -- & -- & -- & 8.48\\
\br
\end{tabular}
\\ $^a$ Central force nonlocal, tensor force local.\\
$^b$ Corrected for meson-exchange currents and relativity.\\
$^c$ Reference~\cite{RK90}.
\label{tab_pots}
\end{table}

It is now of interest to know by how much
the local approximation changes the original amplitude.
This is demonstrated in figure~\ref{fig_pot3sd1}, where the half off-shell
$^3S_1$--$^3D_1$ potential, 
which can be produced only by tensor forces,
is shown.
The on-shell momentum $q'$ is held fixed at 265 MeV/c
(equivalent to 150 MeV laboratory energy),
while the off-shell momentum $q$ runs from zero
to 2000 MeV/c.
The on-shell point ($q=265$ MeV/c) is marked by a solid dot.
The solid curve is the CD-Bonn potential which contains
the full, nonlocal OPE amplitude equation~(\ref{eq_operel}).
When the static/local approximation, equation~(\ref{eq_opeq}), is made,
the dashed curve is obtained.
When this approximation is also used for the one-$\rho$
exchange, the dotted curve results.
It is clearly seen that the static/local approximation
substantially increases the tensor force off-shell.
Certainly, we are not dealing here with negligible effects,
and the local approximation is obviously not a good one.

\begin{figure}
\vspace{0.5cm}
\hspace*{1.80cm}
\epsfig{file=fig_pot1s0.ps,width=5.4cm}

\vspace*{-5.25cm}
\hspace*{7.30cm}
\epsfig{file=fig_pot3s1.ps,width=5.4cm}

\vspace*{1.0cm}
\hspace*{1.75cm}
\epsfig{file=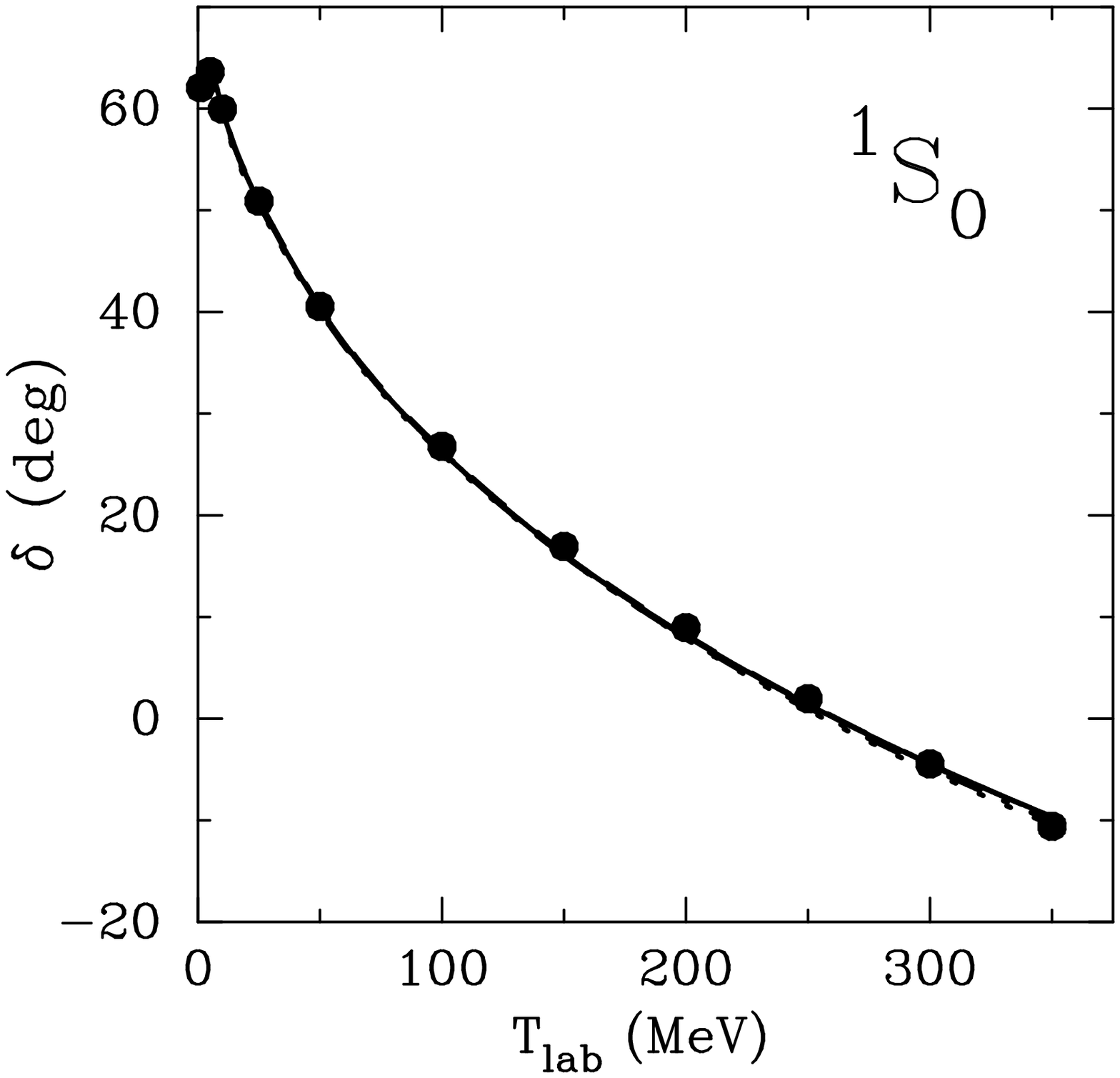,width=8cm}

\vspace*{-6.1cm}
\hspace*{7.25cm}
\epsfig{file=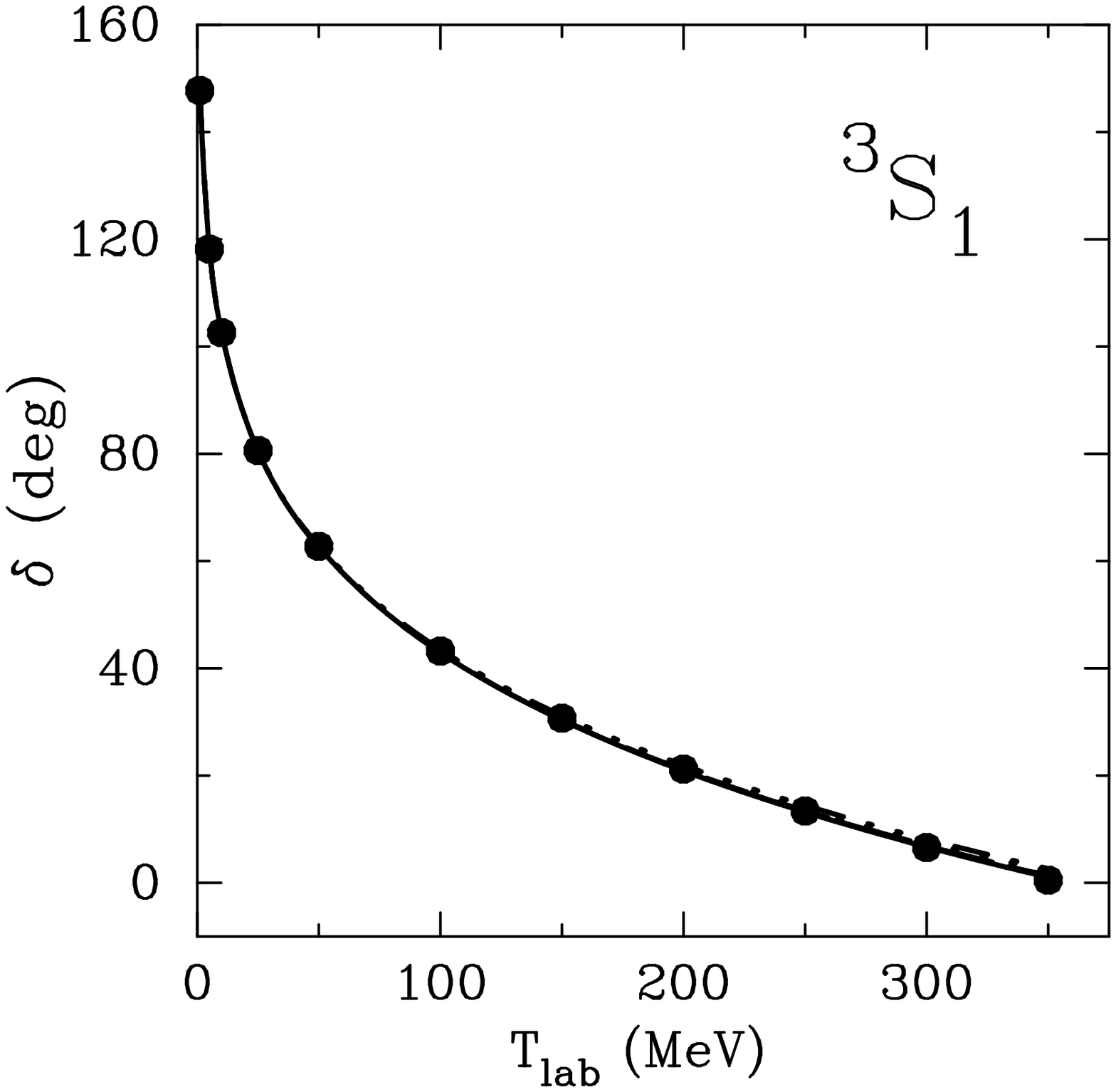,width=8cm}

\vspace*{-1.5cm}
\caption{{\bf Upper part:}
Matrix elements $V(q',q)$ of the $^1S_0$ and $^3S_1$
potentials for the 
CD-Bonn (solid line),
Nijm-I (dashed), Nijm-II (dash-dot), Argonne $V_{18}$
(dash-triple-dot) and Reid93 (dotted) potentials.
The diagonal matrix elements with $q'=q=265$ MeV/c (equivalent
to $T_{lab}=150$ MeV) are marked by a solid dot.
The corresponding matrix element of the scattering $K$-matrix is marked 
by the star.
{\bf Lower part:}
Predictions for the $np$ phase shifts in the $^1S_0$
and $^3S_1$ state by the five potentials.
The five curves are essentially indistinguishable. The solid dots represent
the Nijmegen multi-energy $np$ analysis~\cite{Sto93}.}
\label{fig_pots}
\end{figure}

Even though the spirit of the new generation of potentials is
more sophisticated, only the CD-Bonn potential uses
the full, original, nonlocal Feynman amplitude for OPE, 
equation~(\ref{eq_operel}),
while all other potentials still apply the local approximation,
equations~(\ref{eq_opeq})
and (\ref{eq_oper}).
As a consequence of this, the CD-Bonn potential
has a weaker tensor force as compared to all other potentials.
This is reflected in the predicted D-state probability of
the deuteron, $P_D$, which is due to the nuclear tensor force.
While CD-Bonn predicts $P_D=4.85$\%, the other potentials
yield $P_D= 5.7(1)$\% (cf.\ table~\ref{tab_pots}).
These differences in the strength of the tensor force lead to
considerable differences in nuclear structure predictions.
An indication of this is given in table~\ref{tab_pots}:
The CD-Bonn potentials predicts 8.00 MeV for the triton binding
energy, while the local potentials predict only 7.62 MeV.
More discussion of this aspect can be found in 
references~\cite{Mac89,Mac98}.

The OPE contribution to the nuclear force essentially takes care of the
long-range interaction and the tensor force.
In addition to this, all models must describe the intermediate
and short range interaction, for which very different
approaches are taken.
The CD-Bonn includes (besides the pion)
the vector mesons $\rho (769)$ and $\omega (783)$, 
and two scalar-isoscalar bosons, $\sigma$, 
using the full, nonlocal Feynman amplitudes, 
equation~(\ref{eq_amp}), for their exchanges. 
Thus, all components of the CD-Bonn are nonlocal and the off-shell
behavior is the original one that is determined from
relativistic field theory.

The models Nijm-I and Nijm-II are based upon the Nijmegen78
potential~\cite{NRS78}
which is constructed from approximate OBE amplitudes.
Whereas the Nijm-II uses the totally local approximations
for all OBE contributions, the Nijm-I keeps some nonlocal
terms in the central force component (but the Nijm-I
tensor force is totally local).
Nonlocalities in the central force have only a very
moderate impact on nuclear structure as compared to
nonlocalities in the tensor force. Thus, if for some reason one
wants to keep only some of the original nonlocalities 
in the nuclear force and not all of them,
then it would be more important to keep the tensor force nonlocalities.

The Reid93~\cite{Sto94} and Argonne $V_{18}$~\cite{WSS95} potentials do not 
use meson-exchange for 
intermediate and short range; instead, a phenomenological parametrization
is chosen.
The Argonne $V_{18}$ uses local functions
of Woods-Saxon type, 
while Reid93 applies local Yukawas of multiples
of the pion mass, similar to the original Reid potential
of 1968~\cite{Rei68}.
At very short distances, the potentials are regularized 
either by 
exponential ($V_{18}$, Nijm-I, Nijm-II) or by dipole (Reid93)
form factors (which are all local functions).

In figure~\ref{fig_pots}, the five high-precision potentials
(in momentum space) and their phase shift predictions
are shown,
for the $^1S_0$ and $^3S_1$ states.
While the phase shift predictions are indistinguishable,
the potentials differ widely---due to the theoretical and mathematical
differences discussed. Note that NN potentials differ the most in
$S$-waves and converge with increasing $L$ (where $L$ denotes the
total orbital angular momentum of the two-nucleon system).

\subsection{Charge dependence}
All new potentials are charge-dependent which is
essential for obtaining a good $\chi^2$ for the $pp$ and $np$ data. 
Thus, each potential
comes in three variants: $pp$, $np$, and $nn$.

All potentials include the CIB effect from OPE.
However, as discussed in section 2.2,
pion mass splitting creates further CIB effects through
the diagrams of $2\pi$ exchange and other two-boson exchange diagrams
that involve pions. 
Another source of CIB is irreducible $\pi\gamma$ exchange. 
Recently, these contributions have been
evaluated in the framework of chiral perturbation theory
by van Kolck {\it et al.}~\cite{Kol98}. 
In $L>0$ states, the size of this contribution is typically
the same as the CIB effect from TBE.
Thus, TBE and $\pi\gamma$ create
sizable CIB effects in states with $L>0$. 
Therefore, a thoroughly constructed, modern, charge-dependent
NN potential should include them.
The NN potentials~\cite{Sto94,WSS95} ignore
these contributions while the latest CD-Bonn update~\cite{Mac00} 
incorporates them.

A similar comment can be made about CSB.
Most potentials include only the most trivial effects from
nucleon mass splitting, namely the effect on the kinetic
energy and on the OBE diagrams. However, as discussed in
section 2.1.2, there are relatively large contributions
from TBE that fully explain the CSB scattering length
difference.
Because of the outstanding importance of the CSB effect from TBE, 
it should be included in NN force models 
(and, therefore, it has been incorporated
in the latest update of the CD-Bonn potential~\cite{Mac00}). 
To have distinct $pp$ and $nn$ potentials 
is important for addressing several interesting issues
in nuclear physics, like, the $^3$H-$^3$He binding energy difference
and the Nolen-Schiffer (NS) anomaly~\cite{NS69}
regarding the energies of neighboring mirror nuclei. 
Potentials that do not include any CSB have no chance to ever
explain these phenomena.
Some potentials that include CSB focus on the $^1S_0$ state only, since 
this is where the most reliable empirical information is.
However, even this is not good enough.
A recent study~\cite{MPM99} has shown that
the CSB in partial waves with $L>0$
as derived from the Bonn model
is crucial for a quantitative explanation of the NS anomaly.

\subsection{Extrapolating low-energy potentials towards higher energies}
NN potentials designed for nuclear structure purposes are typically
fitted to the NN scattering data up to pion production
threshold or slightly beyond (e.~g., 350 MeV).
A very basic reason for this is that a real potential cannot
describe the inelasticities of particle production.
On the other hand, nuclear structure calculations are
probably sensitive to the properties of a potential
above 350 MeV. For example, the Brueckner $G$-matrix,
which is a crucial quantity in many microscopic approaches
to nuclear structure, is the solution of the integral
equation,
\begin{equation}
G({\bf q'}, {\bf q}) = V({\bf q'}, {\bf q}) -
\int d^3k V({\bf q'}, {\bf k}) \frac{M^\star Q}{k^2-q^2}
G({\bf k}, {\bf q}) 
\end{equation}
(where $M^\star$ denotes the effective nucleon mass and
$Q$ the Pauli projector).
Notice that the potential $V$ is involved in this equation for all
momenta from zero to infinity, on- and off-shell.
Now, it may very well be true that,
as the momenta increase,
their importance decreases 
(due to the short-range repulsion of the nuclear force
and the associated
short-range suppression of the nuclear wave function).
However, it is also true that
the impact of the potential does not suddenly drop to zero
as soon as the momenta involved become larger than the
equivalent of 350 MeV lab.\ energy.
Thus, there are good arguments why
NN potentials should
extrapolate in a reasonable way towards higher energies.

\begin{figure}
\vspace*{-2.5cm}
\hspace*{-2.5cm}
\epsfig{figure=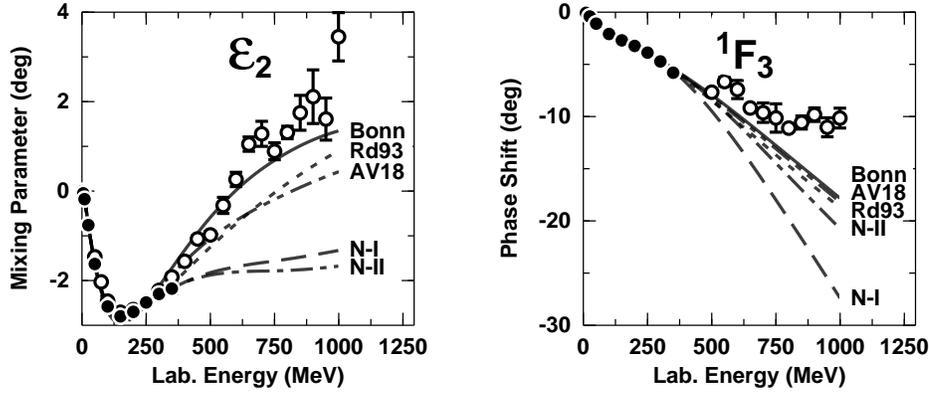,width=19.0cm}
\vspace*{-15.0cm}
\caption{The $\epsilon_2$ mixing parameter and the $^1F_3$ phase shift 
up to 1000 MeV lab.\ energy for various potentials
as denoted (N-I and N-II refer to the Nijmegen potentials). 
Solid dots represent the Nijmegen PSA~\protect\cite{Sto93}
and open circles the VPI/GWU analysis SM99~\protect\cite{SAID}.}
\label{fig_extra}
\end{figure}

We have investigated this issue and found good and bad
news. The good news is that most potentials reproduce 
in most partial waves the
NN phase shifts up to about 1000 MeV amazingly well.
The bad news is that there are some singular cases
where the reproduction of phase parameters for higher energies
is disturbingly bad.
The two most notorious cases are shown in figure~\ref{fig_extra}.
Above 350 MeV, the $\epsilon_2$ mixing parameter is substantially
underpredicted by both Nijmegen potentials (N-I and N-II).
The reason for this is that, 
for $\epsilon_2$, 
both potentials follow 
very closely 
the Nijmegen PSA~\cite{Sto93} (solid dots in figure~\ref{fig_extra}) 
up to 350 MeV. 
Thus, these potentials are 
faithfull extrapolations of the Nijmegen PSA to higher energies.
Since this extrapolation is wrong, the suspicion is that
the Nijmegen PSA has a wrong trend in the energy range
250-350 MeV. New data on $pp$ spin transfer coefficients~\cite{Wi99}
in the energy range 300-500 MeV could resolve the issue.

A similar problem occurs in $^1F_3$ (figure~\ref{fig_extra}).
Here, the dashed curve (N-I) is the extrapolation of the
Nijmegen PSA, indicating that the analysis 
may have the wrong trend
in the energy range 200-350 MeV. 

We note that, in the two channels discussed, the inelasticity has little
impact on the phase parameters shown and would not fix the problems.

The moral is that one should not follow just one analysis,
particularly, if that analysis is severely limited in its energy range. 
It is important to also keep the broad picture in mind.

\section{The theory of nuclear forces: future directions}

\subsection{Critical summary of current status}

During the past decade or so, the research on the NN interaction
has proceeded essentially along two lines.
There was the phenomenological line which has produced
the high-precision, charge-dependent
NN potentials~\cite{Sto94,WSS95,MSS96,Mac00}.
This was practical work, necessary to provide reliable 
input for exact few-body calculations and
nuclear many-body theory. 

The goal of the second line of research was to approach the 
problem on a more fundamental level.
Since about 1980, we have seen many efforts to derive the nuclear
force from the underlying theory of strong interactions, 
quantum chromodynamics (QCD).
Due to its nonperturbative character in the low-energy regime,
QCD cannot be solved exactly for the problem under consideration.
Therefore, so-called QCD-related or QCD-inspired models have been 
developed, like, Skyrmion or Soliton models~\cite{skyrme} 
and constituent quark cluster models.
Among the quark models, one may distinguish between two types:
the hybrid models~\cite{qclust} that include meson exchange
and the quark delocalization and color screening models~\cite{QDCSM} that 
do not need (and do not include)
meson exchange to create the intermediate-range
attraction of the nuclear force.
The success of all these efforts based upon QCD modelling is mixed. 
The qualitative features of the nuclear force are, in general,
predicted correctly, but none of the models
is sufficiently quantitative such that it would make sense
to apply it in miscroscopic nuclear structure calculations.

In summary, one problem of the current status in the field 
is that quantitative models for the nuclear force
have only a poor theoretical background, while theory based models
yield only poor results.
This discrepancy between theory and practice has become rather larger 
than smaller, in the course of the 1990s.  
Another problem is that the `theory based models' are not strictly derived from
QCD, they are modeled after QCD---often with handwoven arguments.
Thus, one may argue that these models are not any better than the traditional
meson-exchange models (that are nowadays perceived as phenomenology).
The purpose of physics is to explain nature in fundamental terms.
The two trends just discussed are moving us away from this aim, which is
reason for serious concern.

Therefore, the main goal of future research on the nuclear force must be to
overcome the above discrepancies.
To achieve this goal, {\it we need a basic theory
that is amenable to calculation and yields quantitative results.}

\subsection{The effective field theory concept}

In recent years, the concept of effective field theories (EFT) 
has drawn considerable attention in particle and nuclear 
physics~\cite{Geo93,Eck95,Kap95}.
The notion of {\it effective} field theories may suggest a difference
to {\it fundamental} field theories.
However, it is quite likely that all field theories (including those
that we perceive presently as fundamental) are effective in the sense
that they are low-energy approximations to some `higher' theory.

The basis of the EFT concept is the recognition of different
energy scales in nature.
Each energy level has its characteristic degrees of freedom.
As the energy increases and smaller distances are probed, new
degrees of freedom become relevant and must be included.
Conversely, when the energy drops, some degrees of freedom 
become irrelevant and are frozen out.

To model the low-energy theory, one relies on a famous `folk theorem' by
Weinberg~\cite{Wei79,Wei97} which states:
\begin{quote}
If one writes down the most general possible Langrangian, including {\it all}
terms consistent with assumed symmetry principles,
and then calculates matrix elements with this Langrangian to any given order of
perturbation theory, the result will simply be the most general possible 
S-matrix consistent with analyticity, perturbative unitarity,
cluster decomposition, and the assumed symmetry principles.
\end{quote}
The essential point of an effective field theory is that we are not allowed to make
any assumption of simplicity about the Lagrangian and, consequently,
we are not allowed to assume renormalizability. 
The Langrangian must include all possible terms, because this completeness
guarantees that the effective theory is indeed the low-energy
limit of the underlying theory.
Now, this implies that we are faced with an infinite set of interactions.
To make the theory managable, we need to organize a perturbation expansion.
Then, up to a certain order in this expansion, the number of terms that
contribute is finite and the theory will yield a well-defined result.

In strong interactions, the transition from the `fundamental' to the
effective level happens through a phase transition that takes place
around $\Lambda_{QCD}\approx 1$ GeV 
via the spontaneous breaking of chiral symmetry 
which generates pseudoscalar Goldstone bosons.
Therefore, at low energies 
($E<\Lambda_{QCD}$), 
the relevant degrees of freedom
are not quarks and gluons, but pseudoscalar mesons and other hadrons.
Approximate chiral symmetry is reflected in the smallness of the masses
of the pseudoscalar mesons.
The effective theory that describes this scenario
is known as chiral perturbation theory ($\chi$PT)~\cite{Eck95,Leu94,BKM95}.

If we believe in the basic ideas of EFT, then, at low energies, $\chi$PT
is as fundamental as QCD at high energies.
Moreover, due to its perturbative arrangement, $\chi$PT can be calculated:
order by order. So, here we may have what we are asking for:
{\it a basic theory that is amenable to calculation.}
Therefore, $\chi$PT has the potential to overcome the discrepancy between
theory and practice that has beset the theoretical research on the 
nuclear force for so many years.

\subsection{Chiral perturbation theory and nuclear forces}

The idea to derive nuclear forces from chiral effective Lagrangians
is not new. A program of this kind was started some 10 years ago by 
Weinberg~\cite{Wei90,Wei92a}, Ord\'o\~nez~\cite{OK92},
and van Kolck~\cite{Kol93,Kol94,Kol99} which produced the
first chiral NN potential by Ord\'o\~nez, Ray, and van Kolck~\cite{ORK94}.

After the program was initiated, considerable activity
ensued~\cite{FST97,CPS92,Bir94,RR97,Par99,KSW98,Fri97,KBW97,Coh97,FMS99,RS99,BHK99}.
For a recent review, see Ref.~\cite{Bea00}.
Even though all authors start from chiral effective Langrangians,
there are differences in the details.
There is, for example, the KSW scheme~\cite{KSW98}
in which the amplitude of interest is calculated 
perturbatively.
On the other hand, Weinberg proposed to use
$\chi$PT for computing the NN potential
which consists of irreducible diagrams.
The $S$-matrix is then generated by the Schr\"odinger equation.
The first comprehensive work using the Weinberg scheme
was done by Ord\'o\~nez, Ray, and van Kolck~\cite{OK92,Kol93,ORK94}
who applied time-ordered (`old-fashioned') perturbation theory
to calculate the irreducible diagrams that define the potential.
This potential posesses one unpleasant property, namely,
it is energy dependent.
One can avoid this problem by using the method of unitary
transformations, a method that was pioneered by Okubo~\cite{Oku54}.
The Okubo tranformation is applied in the recent work by
Epelbaum, Gl\"ockle, and Mei\ss ner~\cite{EGM98,EGM00}
who construct chiral NN potentials in leading order (LO) of
$\chi$PT, next-to-leading order (NLO), and next-to-next-to-leading
order (NNLO).
A systematic improvement in the ability of the model to reproduce
the NN data is obseerved when stepping up the orders of the
chiral expansion. The NNLO potential of 
Epelbaum {\it et al.}~\cite{EGM00} describes the $np$ phase shifts
well up to about 100 MeV; above this energy there are 
discrepancies in some partial waves. This most recent chiral NN potential
represents great progress as compared to earlier ones, however,
for meaningfull applications in microscopic nuclear structure,
futher quantitative improvements are necessary.

There is one particularly attractive aspect to the $\chi$PT
approach in regard to those nuclear structure applications.
If, in the traditional approach,
one wants to reproduce, e.~g., the experimental binding energies of
the triton, the alpha particle or other nuclei,
one complements the NN potential with a (phenomenological)
three-nucleon force (3NF)~\cite{NKG00}. Since different NN potentials leave
different discrepancies to experiment (cf.\ table~\ref{tab_pots}), 
the 3NF is adjusted
from potential to potential. 
From a more fundamental point of view, this proceedure is
very unsatisfactory, since it lacks any underlying systematics.
However, within the framework of traditional
meson theory, there is nothing else you can do,
because there is no {\it a priori} connection between the off-shell
NN potential and the existence of certain many-body forces.

In the framework of $\chi$PT, there is this connection
from the outset.  In each order of $\chi$PT,
the two-nucleon force is well-defined on- and off-shell
{\it and} it is also well-defined which 3NF terms 
occur in that order.
At least that's how it should work `in theory'.
How it works out in practise remains to be seen.

\section{NN scattering at intermediate and high energies}

Even though the focus of this review is on the NN interaction at
low energy (below pion production threshold),
we like to give an indication of the exciting developments
at higher energies to ensure that the reader gets
an impression of the broader picture.

In the 1990's,
a new experimental effort to measure $pp$
scattering observables in the range 0.5 to 2.5 GeV
of projectile energy was started.
The experiments are conducted by the EDDA
group~\cite{EDDA1,EDDA2} at the cooled proton 
synchrotron (COSY) at J\"ulich, Germany, and use an internal target.
The outstanding features of
this new generation of experiments 
are high precision (high statistics) and small energy steps.
Other experiments on NN observables at intermediate  and high energies
have been performed at SATURN II (Saclay, France)~\cite{All99}
and at facilities around the world; for a complete listing, see
Ref.~\cite{Arn97}.
Moreover, there are the $pp$ analyzing powers and spin correlation
parameters, $C_{NN}$, that were measured up to lab.\ energies
of 11.75 GeV by Alan Krisch and coworkers 
some 20 years ago at the Argonne Zero-Gradient-Synchrotron 
(ZGS)~\cite{FK81}. The latter data were never explained
in a really satisfatory way in spite of some efforts~\cite{BT88,HR90}.
During the past decade, 
theory has shown little interest in NN scattering
at energies of 1-10 GeV
and only a few papers can be cited~\cite{Neu91,LLA93,Ger98,Koc99,KL99,GL00}.
The reason for the neglect is probably that these energies are
too high for traditional nuclear physicists and too low for
high energy physicists; so, this range is the stepchild of both
professions.
But this is what makes this window particularly interesting:
the energies are too high for $\chi$PT and too low for perturbative
QCD.
This fact implies that we have to find an appropriate
phenomenology which calls for some phantasy and creativity.
Relativistic, chiral meson models that include heavy mesons [like,
$\rho(770)$ and $\omega(782)$] and nucleon resonances
[like, the $\Delta(1232)$ isobar] are an obvious choice.
In the language of EFT, these models can be justified
on the basis of `resonance saturation'. It will also be interesting
to attempt matching of these high energy models with
the $\chi$PT model discussed in the previous section.

It is well-known that relativistic meson models
work satisfactorily up to about 1.2 GeV~\cite{KS80,Lee83,FT84,HM87,Els88};
for a summary and critical discussion see 
section 7 of Ref.~\cite{Mac89}.
However, at energies above 1.5 GeV, characteristic
problems occur, some of which are~\cite{Mac99}:
\begin{itemize}
\item
The predicted elastic NN cross sections are too large and grow with energy
while experimentally they drop.
\item
The $pp$ analyzing powers are predicted too large and for other
spin observables (like, $C_{NN}$) even the sign is predicted
wrong.
\end{itemize}

The first problem listed above is well-known since the late 1950's.
The amplitude produced by vector-meson exchange is proportional to
$s/(t-m_v^2)$
(with $m_v$ the vector-meson mass and $s,t$ the usual Mandelstam
variables) which causes the elastic cross sections to rise with energy.
Because of this problem, Regge theory~\cite{Col77,Per74,Can89,Mat94} 
was invented.
Concerning the second problem listed above, we do not have a
clue at this time. When the vector-meson contribution is
`phased out' such as to produce the correct elastic cross sections,
then the problem with the analyzing powers persists, which is
not what we expected.
In any case, this energy region offers a wealth of good and unexplained
data and a great diversity of potentially appropriate models,
since we are truely at the intersection of nuclear and
particle physics.

\section{Summary and outlook}

In the 1990's, we made major progress 
in our grasp on the nuclear force. 
NN data of unparalleled precision were produced, particularly
at TUNL~\cite{Wi95,Ra99},
IUCF~\cite{Wi99,Ra98,Lo00,Ha97,Pr98}, 
and COSY~\cite{EDDA1,EDDA2}.
The art of NN phase shift analysis advanced substantially~\cite{Sto93}.
Based upon this empirical progress, NN potentials of 
unprecedented accuracy
($\chi^2$/datum $\approx 1$) were contructed~\cite{Sto94,WSS95,MSS96,Mac00}.
As a consequence of this, exact few-body calculations and miscroscopic
nuclear many-body theory can now be based upon input that is more
reliable than ever.

In spite of all this good news, there are also several questions concerning the
NN interaction that remain open and future research should continue to address
them.
Among the more technical problems is the neutron-neutron
 scattering length where experiments are still in contradiction.
There are problems with the determination of a precise
value for the $\pi NN$ coupling constant and, not unrelated,
the huge and expensive database of neutron-proton
differential cross sections~\cite{Bo78,Ro92,Ra98,Fr99}
is inconsistent.

On the theoretical side, we are still lacking a derivation of the nuclear
force that is based upon theory (in the true sense of the word)
and produces a quantitative NN potential. Moreover, our understanding
of charge-dependence of the NN interaction is still incomplete,
since we are not able to explain 25\% of the charge-dependence of 
the $^1S_0$ scattering length.

\ack
This work was supported in part by the U.S.\ National Science
Foundation under Grant No.\ PHY-9603097.


\section*{References}
\begin{thebibliography}{999}
\bibitem{Cha32} Chadwick J 1932 {\it Proc. Roy. Soc. (London)} {\bf A136} 692
\bibitem{Ros48} Rosenfeld L 1948 {\it Nuclear Forces} (Amsterdam: North-Holland)
\nonum
Eder G 1965 {\it Kernkr\"afte} (Karlsruhe: Braun)
\bibitem{Yuk35} Yukawa H 1935 {\it Proc. Phys. Math. Soc. Japan} {\bf 17} 48
\bibitem{GT57} Gammel J L and Thaler R M 1957 {\it Phys. Rev.}
{\bf 107} 291 and 1339 
\bibitem{Mac89} Machleidt R 1989 {\it Adv. Nucl. Phys.} {\bf 19} 189
\bibitem{ML94} 
Machleidt R and Li G Q 1994 {\it Phys. Rep.} {\bf 242} 5
\bibitem{SAT89} Slaus I, Akaishi Y and Tanaka H 1989 {\it Phys. Rep.}
{\bf 173} 257
\bibitem{MNS90} Miller G A, Nefkens M K, 
and Slaus I 1990 {\it Phys. Rep.} {\bf 194} 1 
\bibitem{MO95} Miller G A and van Oers W H T 1995 {\it Symmetries
and Fundamental Interactions in Nuclei}
Haxton W C and Henley E M (ed) (Singapore: World Scientific)
p~127
\bibitem{Glash} 
Glashow D W {\it et al} 1986 {\it Radiation Effects} {\bf 94} 913
\nonum Glashow D W 1992 {\it LANL Report} LA-UR-92
\bibitem{How98} Howell C R {\it et al} 1998 {\it Phys. Lett.} B {\bf 444} 252
\bibitem{Gon99} Gonz\'alez Trotter D E {\it et al} 1999
{\it Phys. Rev. Lett.} {\bf 83} 3788
\bibitem{Gab79} 
Gabioud B {\it et al} 1979 
{\it Phys. Rev. Lett.} {\bf 42} 1508 
\nonum
Gabioud B {\it et al} 1981 
{\it Phys. Lett.} B {\bf 103} 9 
\nonum
Gabioud B {\it et al} 1984 
{\it Nucl. Phys.} A {\bf 420} 496
\bibitem{Sch87} Schori O \etal 1987 {\it Phys. Rev.} C {\bf 35} 2252
\bibitem{Sla82} Slaus I {\it et al} 1982 {\it Phys. Rev. Lett.}
{\bf 48} 993 
\bibitem{Glo96} Gl\"ockle W {\it et al} 1996 {\it Phys. Rep.}
{\bf 274} 107
\bibitem{Set96} Setze H R \etal 1996 {\it Phys. Lett.} B {\bf 388} 229
\bibitem{Tor96} 
Tornow W 
{\it et al} 1996 {\it Few body systems} {\bf 21} 97
\nonum 
Tornow W 2000 Private communication
\bibitem{Slo84} Slobodrian R J \etal 1984 {\it Phys. Lett.} B {\bf 135} 17 
\bibitem{Huh00} Huhn V \etal 2000 {\it Phys. Rev. Lett.} {\bf 85} 1190
\nonum Huhn V \etal 2001 {\it Phys. Rev.} C {\bf 63} 014003
\bibitem{Tip00} Tippens W B {\it et al} 2000 {\it Phys. Rev.}
C submitted
\nonum Batinic M {\it et al} 1997
{\it Proc.\ 15th Intl Conf on Few Body Problems in Physics} Book of
Contributions p~419
\nonum A. Marusic 1997 Ph.D. thesis Univ of Zagreb 
\bibitem{Bat98}
Batinic M {\it et al} 1998 Physica Scripta {\bf 58} 15, 
and references therein 
\bibitem{LM98a} Li G Q and Machleidt R 1998 {\it Phys. Rev.} C {\bf 58} 1393
\bibitem{MHE87} Machleidt R, Holinde K, and Elster C 1987 {\it Phys. Rep.} {\bf 
149} 1
\bibitem{CN96} Coon S A and Niskanen J A 1996 {\it Phys. Rev.} C {\bf 53} 1154
\bibitem{Mut99} M\"uther H, Polls A, and Machleidt R 1999 {\it Phys. Lett.} B
{\bf 445} 259
\bibitem{GHT92} Goldman T, Henderson J A, and Thomas A W 1992 {\it Few-Body
Systems} {\bf 12} 193 
\bibitem{PW93} Piekarewicz J and Williams A G 1993 {\it Phys. Rev.} C {\bf 47}
2462
\bibitem{KTW93} Krein G, Thomas A W, and Williams A G 1993
{\it Phys. Lett.} {\bf B317} 293 
\bibitem{Con94} O'Connell H B, Pearce B C, Thomas A W, 
and Williams A G 1994 {\it Phys. Lett.} {\bf B336} 1
\bibitem{CMR97} Coon S A, McKellar B H J, and Rawlinson A A 
1997
{\it Intersections between Nuclear and Particle Physics}
AIP Conf.\ Proc.\ {\bf 412} edited by T. W. Donelly (Woodbury, N.Y.)
p~368
\bibitem{Con97} O'Connell H B, Pearce B C, Thomas A W, 
and Williams A G 1997
{\it Prog. Part. Nucl. Phys.} {\bf 39} 201 
\bibitem{CS00} Coon S A and Scadron M D 2000
{\it Charge symmetry breaking via $\Delta I = 1$ group theory or by the
$u$-$d$ quark mass difference and direct photon exchange},
Proc.\ XXIII Symposium on Nuclear
Physics, Oaxtepec, Mexico, January 2000,
Revista Mexicana de Fisica, to be published
\bibitem{Kol98} van Kolck U, Rentmeester M C M, Friar J L,
Goldman T, and de Swart J J 1998 {\it Phys. Rev. Lett.} {\bf 80} 4386
\bibitem{KMS84} Houk T L 1971 {\it Phys. Rev.} C {\bf 3} 1886
\nonum 
         Dilg W 1975 {\it Phys. Rev.} C {\bf 11} 103
\nonum 
         Koester L and Nistler W 1975 {\it Z. Physik} {\bf A272} 189 
\nonum 
         Klarsfeld S, Martorell J, and Sprung D W L 1984
         {\it J. Phys. G: Nucl. Phys.} {\bf 10} 165
\bibitem{LM98b} Li G Q and Machleidt R 1998 
{\it Phys. Rev.} C {\bf 58} 3153
\bibitem{PDG98} Particle Data Group 2000 {\it Eur. Phys. J.} C {\bf 15} 1
\bibitem{EM83} Ericson T E O and Miller G A 1983 
{\it Phys. Lett.} {\bf 132B} 32
\bibitem{PL70} Partovi M H  and Lomon E L 1970 {\it Phys. Rev.} D {\bf 2}
1999 
\bibitem{Ban75} Banerjee M K 1975 {\it Electromagnetic Interactions
of Nucleons} University of Maryland Technical Report No.\ 75-050
\bibitem{Che75} Chemtob M 1975 {\it Interaction Studies in Nuclei} edited by
Jochim H and Ziegler B (Amsterdam: North-Holland) p~487
\bibitem{Dum83} Dumbrajs O \etal 1983
{\it Nucl. Phys.} {\bf B216} 277
\bibitem{BCC73} Bugg D V, Carter A A and Carter J R 1973
{\it Phys. Lett.} {\bf B44} 278 
\bibitem{KP80} Koch R and Pietarinen E 1980 {\it Nucl. Phys.} {\bf A336} 331
\bibitem{Kro81} Kroll P 1981 {\it Phenomenological Analysis of
Nucleon-Nucleon Scattering}, Physics Data Vol.~22-1, H. Behrens and
G. Ebel, eds. (Karlsruhe: Fachinformationszentrum)
\bibitem{Ber87} Bergervoet J R , van Campen P C, Rijken T A
and de Swart J J 1987 {\it Phys. Rev. Lett.} {\bf 59} 2255
\bibitem{Ber90} Bergervoet J R, van Campen P C, Klomp R A M,
de Kok J L, Rijken T A, Stoks V G J
and de Swart J J 1990 {\it Phys. Rev.} C {\bf 41} 1435
\bibitem{Arn90} Arndt R A, Li Z J, Roper L D  and Workman R L
1990 {\it Phys. Rev. Lett.} {\bf 65} 157
\bibitem{Tim95} 
Timmermans R G E 1997 
{\it Few-Body Systems Suppl.} {\bf 9} 169 
\nonum 
Timmermans R G E 1997 
{\it $\pi N$ Newsletter} {\bf 13} 80
\nonum 
Timmermans R G E 1998 
{\it Nucl. Phys.} {\bf A631} 343c 
\bibitem{STS93} Stoks V, Timmermans R and de Swart J J 1993
{\it Phys. Rev.} C {\bf 47} 512
\bibitem{AWP94} Arndt R A, Workman R L and Pavan M M 1994
{\it Phys. Rev.} C {\bf 49} 2729
\bibitem{ASW94}
Arndt R A, Strakovsky I I and Workman R L 1994
{\it Phys. Rev.} C {\bf 50} 2731
\bibitem{ASW95}
Arndt R A, Strakovsky I I and Workman R L 1995
{\it Phys. Rev.} C {\bf 52} 2246
\bibitem{MB93} Markopoulou-Kalamara F G and Bugg D V 1993
{\it Phys. Lett.} {\bf B318} 565
\bibitem{BM95} Bugg D V and Machleidt R 1995
{\it Phys. Rev.} C {\bf 52} 1203
\bibitem{Upp99} {\it Proc. Workshop on Critical Issues in the Determination
of the Pion-Nucleon Coupling Constant} (Uppsala, Sweden, June 1999)
2000 Physica Scripta {\bf T87} 1
\bibitem {Eri95} 
Ericson T E O {\it et al} 1995
{\it Phys. Rev. Lett.} {\bf 75} 1046
\bibitem{Rah98} Rahm J {\it et al} 1998 
{\it Phys. Rev.} C {\bf 57} 1077
\bibitem{RKS98} Rentmeester M C M, Klomp R A M and
de Swart J J 1998 
{\it Phys. Rev. Lett.} {\bf 81} 5253
\nonum
Ericson T E O {\it et al} 1998
{\it Phys. Rev. Lett.} {\bf 81} 5254 
\bibitem{SRT98} de Swart J J, Rentmeester M C M and
Timmermans R G E 1998 {\it The Status of the Pion-Nucleon Coupling
Constant} nucl-th/9802084
\bibitem{Wi99} Wissink S W \etal 1999 \PRL {\bf 83} 4498
\bibitem{RK90} Rodning N L and Knutsen L D 1990 
{\it Phys. Rev.} C {\bf 41} 898
\bibitem{ER83} Ericson T E O and Rosa-Clot M 1983 
{\it Nucl. Phys.} {\bf A405} 497
\bibitem{Sto93} Stoks V G J, Klomp R A M, Rentmeester M C M
and de Swart J J 1993 {\it Phys. Rev.} C {\bf 48} 792
\bibitem{MSS96} Machleidt R, Sammarruca F and Song Y 1996
{\it Phys. Rev.} C {\bf 53} 1483
\bibitem{Mac00} Machleidt R 2000
{\it The high-precision, charge-dependent, Bonn nucleon-nucleon
potential (CD-Bonn)} 
{\tt arXiv:nucl-th/0006014}
\bibitem{MS91} Machleidt R and Sammarruca F 1991
{\it Phys. Rev. Lett.} {\bf 66} 564
\bibitem{ML93} Machleidt R and Li G Q 1993
{\it $\pi N$ Newsletter} {\bf 9} 37
\bibitem{HP75} H\"ohler G and Pietarinen E 1975
{\it Nucl. Phys.} {\bf B95} 210
\bibitem{Sak69} Sakurai J J 1969
{\it Currents and Mesons}
(Chicago: University of Chicago Press)
\bibitem{BM94} Brown G E and Machleidt R 1994
{\it Phys. Rev.} C {\bf 50} 1731
\bibitem{SAID} SAID, Scattering Analysis Interactive Dial-in facility
by R. A. Arndt, I. I. Strakovsky, and R. L. Workman,
Virginia Polytechnic Institute and State University, 
The George Washington University and Jefferson Lab.;
for information on SAID and the VPI/GWU phase shift analysis, 
see reference~\cite{ASW94}
\bibitem{Bar82} 
Barker M D {\it et  al} 1982
{\it Phys. Rev. Lett.} {\bf 48} 918
\nonum
Barker M D {\it et  al} 1982
{\it Phys. Rev. Lett.} {\bf 49} 1056
\bibitem{Wei92} Weisel G J {\it et al} 1992
{\it Phys. Rev.} C {\bf 46} 1599
\bibitem{Wil84} Wilczynski K {\it et al} 1984
{\it Nucl. Phys.} {\bf A425} 458
\bibitem{MB99} Machleidt R and Banerjee M K 2000
{\it Few-Body Systems} {\bf 28} 139
\bibitem{Hwa98} Hwang W Y P {\it et al} 1998
{\it Phys. Rev.} C {\bf 57} 61
\bibitem{Bor89} Borbely I, Gr\"uebler W, Vuaridel B and
K\"onig V 1989 {\it Nucl. Phys.} {\bf A503} 349
\bibitem{ER84}
Ericson T E O 1984
{\it Comments Nucl. Part. Phys.} {\bf 13} 157
\nonum
Ericson T E O and Rosa-Clot M 1985
{\it Ann. Rev. Nucl. Part. Sci.} {\bf 35} 271
\nonum
Ericson T E O and Weise W 1988
{\it Pions and Nuclei} (Oxford: Clarendon Press)
\bibitem{BB92} Bugg D V and Bryan R A 1992
{\it Nucl. Phys.} {\bf A540} 449
\bibitem{NRS78} Nagels M M, Rijken T A and de Swart J J 1978
{\it Phys. Rev.} D {\bf 17} 768
\bibitem{Do97} Dombrowski H, Khoukaz A and Santo R 1997
{\it Nucl. Phys.} {\bf A619} 97
\bibitem{Kr94} Kretschmer W {\it et al} 1994 
{\it Phys. Lett.} {\bf B328} 5
\bibitem{Ra98} Rathmann R {\it et al} 1998
{\it Phys. Rev.} C {\bf 58} 658
\bibitem{Lo00} 
Lorentz B 1998
Ph.D.\ thesis University of Wisconsin-Madison
\nonum
Lorentz B \etal 2000
{\it Phys. Rev.} C {\bf 61} 054002
\bibitem{Ha97} Haeberli W {\it et al} 1997
{\it Phys. Rev.} C {\bf 55} 597
\bibitem{Pr98} v. Przewoski B {\it et al} 1998
{\it Phys. Rev.} C {\bf 58} 1897
\bibitem{SP98} SAID (reference~\cite{SAID}) solution SP98
\bibitem{WSS95} Wiringa R B, Stoks V G J and Schiavilla R 1995
{\it Phys.\ Rev.} C {\bf 51} 38
\bibitem{Wi95} Wilburn W S, Gould C R, Haase D G, Huffman P R,
Keith C D, Roberson N R and Tornow W 1995
{\it Phys. Rev.} C {\bf 52} 2351
\bibitem{Ra99} Raichle B W, Gould C R, Haase D G, Seely M L,
 Walston J R, Tornow W,  Wilburn W S, Penttil\"a S I
and Hoffman G H 1999 {\it Phys. Rev. Lett.} {\bf 83} 2711
\nonum
 Walston J R, Gould C R, Haase D G, Raichle B W, Seely M L,
Tornow W,  Wilburn W S, Penttil\"a S I
and Hoffman G H 2001 {\it Phys. Rev.} C {\bf 63} 014004
\bibitem{BM97} Buerkle W and Mertens G 1997
{\it Few-Body Systems} {\bf 22} 11
\bibitem{Cl98} Clotten P, Hempen P, Hofenbitzer K, Huhn V, 
Metschulat W, Schwindt M, W\"atzold L, Weber C and von Witsch W 1998
{\it Phys. Rev.} C {\bf 58} 1325
\bibitem{Br96} Bro\v{z} J {\it et al} 1996 {\it Z. Physik} {\bf A354} 401
\bibitem{Br97} Bro\v{z} J {\it et al} 1997 {\it Z. Physik} {\bf A359} 23
\bibitem{Be97} Benck S, Slypen I, Corcalciuc V and Meulders J P 1997
{\it Nucl. Phys.} {\bf A615} 220
\bibitem{Go94} Goetz J {\it et al} 1994 {\it Nucl. Phys.} {\bf A574} 467
\bibitem{Ra98a} Rahm J {\it et al} 1998 {\it Phys. Rev.} C {\bf 57} 1077
\bibitem{Da96} Davis C A {\it et al} 1996 {\it Phys. Rev.} C {\bf 53} 2052
\bibitem{Fr99} Franz J, R\"ossle E, Schmitt H and Schmitt L 
2000 {\it Physica Scripta} {\bf T87} 14
\nonum Arndt R A 1999 Private communication
\bibitem{Ah98} Ahmidouch A {\it et al} 1998 {\it Eur. Phys. J.} {\bf C2} 627
\bibitem{Ba93} Ball J {\it et al} 1993 {\it Nucl. Phys.} {\bf A559} 489
\bibitem{Ba94} Ball J {\it et al} 1994 {\it Nucl. Phys.} {\bf A574} 697
\bibitem{Bo78} Bonner B E {\it et al} 1978 Phys. Rev. Lett. {\bf 41} 1200
\bibitem{Ro92} R\"onnqvist T {\it et al.} 1992 Phys. Rev. C {\bf 57} 1077
\bibitem{Arn99} Arndt R A 1999 Private communication
\bibitem{Lac80} Lacombe M, Loiseau B, Richard J M, Vinh Mau R, 
C\^{o}t\'{e} J, Pir\`{e}s P and de Tourreil R 1980 {\it Phys. Rev.} C {\bf 21} 861 
\bibitem{Sto94} Stoks V G J, Klomp R A M, Terheggen C P F 
and de Swart J J 1994 {\it Phys.\ Rev.} C {\bf 49} 2950
\bibitem{Mac98} Machleidt R 1999 
{\it Proc.\ Nuclear Structure 98}, Gatlinburg, Tennessee, 1998, AIP Conf.\
Proc.\ {\bf 481}, edited by C. Baktash (Woodbury, N.Y.: AIP) p~3
\bibitem{Rei68} Reid R V 1968 {\it Ann. Phys. (N.Y.)} {\bf 50} 411
\bibitem{NS69} Nolen J A and Schiffer J P 1969
{\it Annu.\ Rev.\ Nucl.\ Sci.} {\bf 19} 471
\bibitem{MPM99}
M\"{u}ther H, Polls A and Machleidt R 1999
Phys. Lett. B {\bf 445} 259
\bibitem{skyrme}
Jackson A, Jackson A D and Pasquier V 1985
{\it Nucl. Phys.} {\bf A432} 567
\nonum
Vinh Mau R, Lacombe M, Loiseau B, Cottingham W N
and Lisboa P 1985 {\it Phys. Lett.} {\bf 150B} 259
\nonum
Walhout T S and Wambach J 1992
{\it Int. J. Mod. Phys.} {\bf E1} 665
\nonum
Pepin S, Stancu F, Koepf W and Wilets L 1996
{\it Phys. Rev.} C {\bf 53} 1368
\bibitem{qclust}
Oka M and Yazaki K 1981 {\it Prog. Theor. Phys.} {\bf 66} 551 and 577
\nonum
Faessler A, Fernandez F, L\"ubeck G, and Shimizu K 1982
{\it Phys. Lett.} B {\bf 112} 201
\nonum
Faessler A, Fernandez F, L\"ubeck G, and Shimizu K 1983 
{\it Nucl. Phys.} {\bf A402} 555
\nonum
Maltman K and Isgur N 1984 {\it Phys. Rev.} D {\bf 29} 952
\nonum
Myhrer F and Wroldsen J 1988 {\it Rev. Mod. Phys.} {\bf 60} 629
\nonum
Takeuchi S, Shimizu K and Yazaki K 1989
{\it Nucl. Phys.} {\bf A504} 777
\nonum
Fernandez F, Valcarce A, Straub U and Faessler A 1993
{\it J. Phys. G: Nucl. Part. Phys.} {\bf 19} 2013
\nonum
Entem D R, Fernandez F and Valcarce A 2000
{\it Phys. Rev.} C {\bf 62} 034002
\nonum
Shimizu K and Glozman L Y 2000
{\it Phys. Lett.} {\bf B477} 59
\bibitem{QDCSM}
Wang F, Wu G H, Teng L J and Goldman T 1992
{\it Phys. Rev. Lett.} {\bf 69} 2901
\nonum
Wu G H, Teng L J, Ping J L, Wang F and Goldman T 1996
{\it Phys. Rev.} C {\bf 53} 1161
\nonum
Ping J L, Wang F and Goldman T 1999
{\it Nucl. Phys.} {\bf A657} 95
\nonum
Wu G H, Ping J L, Teng L J, Wang F and Goldman T 2000
{\it Nucl. Phys.} {\bf A673} 279
\bibitem{Geo93} Georgi H 1993 {\it Annu. Rev. Nucl. Part. Sci.}
{\bf 43} 209
\bibitem{Eck95} Ecker G 1995 {\it Prog. Part. Nucl. Phys.} {\bf 35} 1
\bibitem{Kap95} Kaplan D B 1995 {\it Effective Field Theories}
{\tt arXiv:nucl-th/9506035}
\bibitem{Wei79} Weinberg S 1979 {\it Physica} {\bf 96A} 327
\bibitem{Wei97} Weinberg S 1996 {\it What is Quantum Field Theory, and
What Did We Think It Is?}, Talk presented at the conference
``Historical Examination and Philosophical Reflections on the Foundations of
Quantum Field Theory'', Boston University, March 1996;
in: {\it Boston 1996, Conceptual foundations and quantum field theory},
p.\ 241; {\tt arXiv:hep-ph/9702027}
\bibitem{Leu94} Leutwyler H 1994 {\it Ann. Phys. (N.Y.)} {\bf 235} 165
\bibitem{BKM95} Bernard V, Kaiser N and Mei\ss ner U G 1995
{\it Int. J. Mod. Phys.} E {\bf 4} 193
\bibitem{Wei90} 
Weinberg S 1990 
{\it Phys. Lett.} B {\bf 251} 288
\nonum
Weinberg S 1991 
{\it Nucl. Phys.} {\bf B363} 3
\bibitem{Wei92a} Weinberg S 1992
{\it Phys. Lett.} B {\bf 295} 114
\bibitem{OK92} Ordonez C and van Kolck U 1992
{\it Phys. Lett.} B {\bf 291} 459
\bibitem{Kol93} van Kolck U 1993 Ph.D. thesis University of Texas,
University of Washington Report DOE/ER/40427-13-N94
\bibitem{Kol94} van Kolck U 1994 {\it Phys. Rev.} C {\bf 49} 2932
\bibitem{Kol99} van Kolck U 1999 {\it Prog. Part. Nucl. Phys.} {\bf 43} 337
\bibitem{ORK94} 
Ordonez C, Ray L and van Kolck U 1994
{\it Phys. Rev. Lett.} {\bf 72} 1982 
\nonum
Ordonez C, Ray L and van Kolck U 1994
{\it Phys. Rev.} C {\bf 53} 2086

\bibitem{FST97} Furnstahl R J, Serot B D and Tang H B 1997
{\it Nucl. Phys.} {\bf A615} 441
\nonum
Rusnak J J and Furnstahl R J 1997 
{\it Nucl. Phys.} {\bf A627} 495
\nonum
Furnstahl R J, Rusnak J J and Serot B D 1997 {\tt nucl-th/9709064}
\nonum
Furnstahl R J, Steel J V and Tirfessa N 1999 {\tt nucl-th/9910048}

\bibitem{CPS92} Celenza L S, Pantziris A and Shakin C M 1992
{\it Phys. Rev.} C {\bf 46} 2213

\bibitem{Bir94}
Birse M C 1994 {\it Phys. Rev.} C {\bf 49} 2212
\nonum
Richardson K G, Birse M C and Mc Govern J A 1997
{\tt hep-ph/9708435}
\nonum
Richardson K G 1999 Ph.D. thesis University of Manchester
{\tt hep-ph/0008118}

\bibitem{RR97} 
da Rocha C A and Robilotta M R 1994 
{\it Phys. Rev.} C {\bf 49} 1818
\nonum
da Rocha C A and Robilotta M R 1995 
{\it Phys. Rev.} C {\bf 52} 531
\nonum
Robilotta M R and da Rocha C A 1997 {\it Nucl. Phys.} {\bf A615} 391
\nonum
Ballot J L, da Rocha C A and Robilotta M R 1998 
{\it Phys. Rev.} C {\bf 57} 1574

\bibitem{Par99} 
Park T S, Min D P and Rho M 1995
{\it Phys. Rev. Lett.} {\bf 74} 4153
\nonum
Park T S, Min D P and Rho M 1996
{\it Nucl. Phys.} {\bf A596} 515
\nonum
Park T S, Min D P and Rho M 1999
{\it Nucl. Phys.} {\bf A646} 83
\nonum
Park T S, Kubodera K, Min D P and Rho M 1998
{\it Phys. Rev.} C {\bf 58} 637

\bibitem{KSW98} 
Kaplan D B, Savage M J and Wise M B 1996 
{\it Nucl. Phys.} {\bf B478} 629
\nonum
Kaplan D B 1997 {\it Nucl. Phys.} {\bf B494} 471
\nonum
Savage M J 1997 {\it Phys. Rev.} C {\bf 55} 2185
\nonum
Kaplan D B, Savage M J and Wise M B 1998 
{\it Nucl. Phys.} {\bf B534} 329
\nonum
Kaplan D B, Savage M J and Wise M B 1998 
{\it Phys. Lett.} {\bf B424} 390
\nonum
Kaplan D B, Savage M J and Wise M B 1999 
{\it Phys. Rev.} C {\bf 59} 617

\bibitem{Fri97} Friar J L 1997 {\it Few-Body Systems} {\bf 22} 161

\bibitem{KBW97} 
Kaiser N, Brockmann R and Weise W 1997 
{\it Nucl. Phys.} {\bf A625} 758
\nonum
Kaiser N, Gerstend\"orfer S and Weise W 1998 
{\it Nucl. Phys.} {\bf A637} 395

\bibitem{Coh97}
Cohen T D 1997 {\it Phys. Rev.} C {\bf 55} 67
\nonum
Phillips D R and Cohen T D 1997 {\it Phys. Lett.} {\bf B390} 7 
\nonum
Phillips D R, Kao C W and Cohen T D 1997
{\it  Phys. Rev.} C {\bf 56} 679
\nonum
Beane S R,  Cohen T D and Phillips D R 1998 
{\it Nucl. Phys.} {\bf A632} 445

\bibitem{FMS99} Fleming S, Mehen T and Stewart I W 1999
{\tt nucl-th/9906056}, {\tt nucl-th/9911001}

\bibitem{RS99} Rupak G and Shoresh N 1999 {\tt nucl-th/9902077}, 
{\tt nucl-th/9906077}

\bibitem{BHK99} 
Bedaque P F, Hammer H W and van Kolck B 1998
{\it Phys. Rev.} C {\bf 58} 641
\nonum
Bedaque P F, Hammer H W and van Kolck B 1999
{\it Nucl. Phys.} {\bf A646} 444

\bibitem{Bea00} 
Beane S R, Bedaque P F, Haxton W C, Phillips D R and Savage M J 2000
{\tt nucl-th/0008064}, this is a chapter prepared for the
{\it Handbook of QCD}

\bibitem{Oku54} Okubo S 1954 {\it Prog. Theor. Phys.} {\bf 12} 603


\bibitem{EGM98} 
Epelbaoum E, Gl\"ockle W and Mei\ss ner U G 1998
{\it Nucl. Phys.} {\bf A637} 107
\bibitem{EGM00} 
Epelbaum E, Gl\"ockle W and Mei\ss ner U G 2000
{\it Nucl. Phys.} {\bf A671} 295

\bibitem{NKG00} Nogga A, Kamada H and Gl\"ockle W 2000 
{\it Phys. Rev. Lett.} {\bf 85} 944

\bibitem{EDDA1} 
Albers D {\it et al} 1997 {\it Phys. Rev. Lett.} {\bf 78} 1652
\bibitem{EDDA2} 
Altmeier M {\it et al} 2000 {\it Phys. Rev. Lett.} {\bf 85} 1819
\bibitem{All99} Allgower C E {\it et al} 1999 {\it Phys. Rev.} C {\bf 60}
054001 and 054002
\bibitem{Arn97}
Arndt R A {\it et al} 1997 {\it Phys. Rev.} C {\bf 56} 3005
\nonum
Arndt R A {\it et al} 2000 {\it Phys. Rev.} C {\bf 62} 034005
\bibitem{FK81} Fernow R C and Krisch A D 1981
{\it Ann. Rev. Nucl. Part. Sci.} {\bf 31} 107
\nonum
Krisch A D 1985 {\it Lectures presented at the School
on High Energy Spin Physics}, Lake Louise, Canada
\bibitem{BT88} Brodsky S J and de Teramond G F 1988
{\it Phys. Rev. Lett.} {\bf 60} 1924
\bibitem{HR90} Hoffmann J and Robson D 1990
{\it Phys. Rev.} C {\bf 42} 1225

\bibitem{Neu91} Neudatchin V G \etal 1991 {\it Phys. Rev.} C {\bf 43} 2499
\bibitem{LLA93}
LaFrance P, Lomon E L and Aw M 1993
{\it Improved coupled
channels and R-matrix models: pp predictions to 1GeV}
MIT preprint CTP\#2133
\bibitem{Ger98} von Geram H V \etal 1998 {\it Phys. Rev.} C {\bf 58} 1948
\bibitem{Koc99} Kochelev N I 1999 {\tt hep-ph/9911480}
\bibitem{KL99} Kharzeev D and Levin E 1999 {\tt hep-ph/9912216}
\bibitem{GL00} Gibbs W R and Loiseau B 2000 in preparation

\bibitem{KS80} 
Kloet W M and Silbar R R 1980
{\it Nucl. Phys.} {\bf A338} 281 and 317
\nonum
Kloet W M and Silbar R R 1981
{\it Nucl. Phys.} {\bf A364} 346
\nonum
Dubach J, Kloet W M and Silbar R R 1982
{\it J. Phys. G} {\bf 8} 475
\nonum
Dubach J, Kloet W M and Silbar R R 1982
{\it Nucl. Phys.} {\bf A466} 573
\bibitem{Lee83}
Lee T S H 1983 
{\it Phys. Rev. Lett.} {\bf 50} 1571
\nonum
Lee T S H 1984 
{\it Phys. Rev.} C {\bf 29} 195
\bibitem{FT84} 
van Faassen E E and Tjon J A 1984
{\it Phys. Rev.} C {\bf 30} 285
\nonum
van Faassen E E and Tjon J A 1986
{\it Phys. Rev.} C {\bf 33} 2105
\bibitem{HM87} ter Haar B and Malfliet R 1987
{\it Phys. Rep.} {\bf 149} 207
\bibitem{Els88}
Elster C, Holinde K, Sch\"{u}tte D and Machleidt R 1988 
{\it Phys. Rev.} C {\bf 38} 1828

\bibitem{Mac99}
Machleidt R 1999
{\it The nucleon-nucleon interaction at intermediate energies},
Proc.\ International Workshop on Intermediate Energy Spin Physics,
J\"ulich, Germany, November 1998, F. Rathmann, W.T.H. van Oers, and
C. Wilkin, eds.\ (Forschungszentrum J\"ulich, J\"ulich) p~169

\bibitem{Col77} Collins P D B  1977 {\it An Introduction to Regge Theory
and High Energy Physics} (Cambridge: Cambridge University Press)
\bibitem{Per74} Perl M L 1974 {\it High Energy Hadron Physics}
(New York: Wiley)
\bibitem{Can89} {\it Regge Theory of Low $p_t$ Hadronic Interactions}
Caneschi L ed. 1989 (Amsterdam: North-Holland)
\bibitem{Mat94} Matthiae G 1994 {\it Rep. Prog. Phys.} {\bf 57} 743

\end{thebibliography}
\end{document}